\documentclass[psfig]{aa}
\input psfig.tex                                                                
                                           
\begin{document}                                                                
\def\et{et al.}                                                                 
\def\egs{erg s$^{-1}$}                                                          
\def\egsc{erg s$^{-1}$ cm$^{-2}$}                                               
\def\msu{M$_{\odot}$\ }                                                         
\def\kms{km s$^{-1}$ }                                                          
\def\kmsM{km s$^{-1}$ Mpc$^{-1}$ }                                              
%
   \title{The ROSAT-ESO Flux Limited X-Ray (REFLEX) Galaxy Cluster 
Survey I: The Construction of the Cluster Sample\thanks{
Based on observations at the European Southern Observatory La Silla,
Chile} }

   \author{H. B\"ohringer\inst{1}, P. Schuecker\inst{1}, L. Guzzo\inst{2}, 
   C.A. Collins\inst{3}, W. Voges\inst{1}, S. Schindler \inst{3}, 
   D.M. Neumann \inst{4}, R. G. Cruddace \inst{5}, S. DeGrandi \inst{2},
   G. Chincarini \inst{2,6}, A. C. Edge \inst{7}, H.T. MacGillivray \inst{8},
   P. Shaver \inst{9}}

   \offprints{H. B\"ohringer}                                                   
                                                                                
   \institute{$^1$ Max-Planck-Institut f\"ur extraterr. Physik,                 
                 D 85740 Garching, Germany\\   
              $^2$ Osservatorio Astronomico di Brera, Merate, Italy\\
              $^3$ Liverpool John Moores University, Liverpool,U.K.\\
              $^4$ CEA Saclay, Service d`Astrophysique, Gif-sur-Yvette,France\\
              $^5$ Naval Research Laboratory, Washington, USA\\
              $^6$ Dipartimento di Fisica, Universita degli Studi di Milano, Italy\\
              $^7$ Physics Department, University of Durham, U.K.\\
              $^8$ Royal Observatory, Edinburgh, U.K.\\
              $^9$ European Southern Observatory, Garching, Germany }

   \date{Received ..... ; accepted .....}                                       
                                                                                
   \maketitle                                                                   
                                                                                
   \markboth {The REFLEX Cluster Survey Sample Construction}{}

\begin{abstract}  
We discuss the construction of an X-ray flux-limited sample of 
galaxy clusters, the REFLEX survey catalogue, to be used for cosmological
studies. This cluster identification and redshift survey was conducted
in the frame of an ESO key programme and is based on                  
candidates selected from the southern part of the ROSAT All-Sky                 
Survey (RASS). For the first cluster candidate selection from 
a flux-limited RASS source list we make use of optical data
from the COSMOS digital catalogue produced from the scans of 
the UK-Schmidt plates. To ensure homogeneity of the sample 
construction process this selection is based only on this one 
well defined optical data base.       
The nature of the candidates selected in this process is subsequently 
checked by a more detailed evaluation of the X-ray and optical source
properties and available literature data. The final identification and
the redshift is then based on optical spectroscopic 
follow-up observations.                                          
                                                                                
In this paper we document the process by which the primary cluster            
candidate catalogue is constructed prior to the optical follow-up               
observations. We describe the reanalysis of the RASS source catalogue           
which enables us to impose a proper flux limit cut to the X-ray source list     
without introducing a severe bias against extended sources. We discuss the     
correlation of the X-ray and optical (COSMOS) data to find galaxy density       
enhancements at the RASS X-ray source positions and the further evaluation of
the nature of these cluster candidates. Based also on the results of the
follow-up observations we provide       
a statistical analysis of the completeness and contamination of the
final cluster sample and show results on the cluster number counts.
The final sample of identified X-ray clusters reaches a flux
limit of $3 \cdot 10^{-12}$ erg s$^{-1}$ cm$^{-2}$ in the 0.1 - 2.4 keV band
and comprises 452 clusters. 
The results imply a completeness of the REFLEX cluster sample
well in excess of 90\% and a contamination by 
non-cluster X-ray sources of less than 9\%, an accuracy sufficient
for the use of this cluster sample for cosmological tests. 
                                                                                
\keywords{Cosmology - Galaxies: clusters - Xrays: galaxies}               
\end{abstract}                                                               
%
                                                                                
\section{Introduction}                                                          
                                                                                
Clusters of galaxies as the largest well defined building blocks of our         
Universe are ideal probes for the study of the cosmic large scale               
structure. Statistical measures of the galaxy cluster population like the       
cluster mass function (e.g. Press \& Schechter 1974, Kaiser 1986,               
Henry \& Arnaud 1991, B\"ohringer \& Wiedenmann 1992, White et al. 1993, 
Bahcall \& Cen 1992, Oukbir and Blanchard 1992, 1997, Eke, Cole \& Frenk 1996,
Viana \& Liddle 1996, Borgani et al. 1999), functions describing         
the spatial distribution as the two-point-correlation function                  
(e.g. Bahcall \& Soneira 1983, Klypin \& Kopilev 1993, 
Bahcall 1988, Lahav et al. 1989,                       
Nichol et al.\ 1992, Dalton et al.\ 1994  Romer et al.\ 1994, Abadi et al.
1998, Borgani et al. 1999, Moscardini et al. 2000, Collins et al. 2000)             
and the density fluctuation power spectrum (e.g. Peacock \& West 1992,
Einasto et al.  1997, Retzlaff et al. 1998, Tadros, Efstathiou, 
\& Dalton 1998, Miller \& Batuski 2000, Schuecker et al. 2000),    
can place                                                                        
very important constraints on the characteristic measures of the matter         
density distribution throughout the Universe and its evolution as a function    
of time. This is due to the fact that the formation of galaxy clusters is       
tightly linked to the formation of the large scale structure in our Universe as 
a whole. That clusters are indeed good tracers of the large-scale 
structure is discussed and demonstrated further in one of the following 
papers by Schuecker et al. (2000).
                                                                                
The crucial step in any of these studies is the careful primary           
selection of the galaxy cluster sample to be used for the cosmological  
investigation. Ideally one would like to select the clusters by their           
mass, thus defining the sample by all clusters above a certain mass limit.      
Alternatively one could also think of a certain size criterion.                 
Such parameters are also the most direct parameters predicted by analytical     
cosmological models (e.g. Press-Schechter 1974 type models) or by N-body        
simulations (e.g. Frenk et al.\ 1990, Cen \& Ostriker 1994, Kofman et al. 1996,
Bryan \& Norman, 1998, Thomas et al. 1998, Frenk et al. 1999).                
Both parameters are not easily obtainable from observations, however.
Thus one has to resort to observable criteria, which should be as closely       
linked to the mass and size of the clusters as possible.                        
                                                                                
Since galaxy clusters were first discovered by their galaxy density       
enhancements, a galaxy richness criterion was the first to be used              
to define and select clusters of galaxies. The first large and very widely      
used compilation was that of George Abell (1958) and  Corwin and Olowin
(Abell, Corwin \& Olowin 1989)                                                  
who's selection criteria were fixed to a minimal galaxy number density       
within a metric radius of 3  $h_{50}^{-1}$ Mpc and a defined magnitude interval.
This catalogue was compiled by eye inspection of the Palomar Sky Survey      
Plates and subsequently of UK Schmidt survey plates. 
Another comprehensive cluster catalogue was compiled visually by Zwicky and
collaborators (Zwicky et al. 1961 -68) with a significantly different
cluster definition.
Later similar cluster    
catalogues were constructed based on machine work using digitized data from     
scans of the optical plates (by COSMOS, see Heydon-Dumbleton et al.\ 1989,      
Lumbsden et al.\ 1992                                                           
and APM, see Maddox et al.\ 1990, Dalton et al.\ 1997) using                   
more objective criteria. Further improvement in the optical cluster searches
was achieved by using multicolor CCD surveys and matched filter techniques
(e.g. Postman et al. 1996, Olsen et al. 1999). But it is still very 
difficult and uncertain to assign a mass to a cluster with a given 
observed richness without comprehensive redshift data.
                                                                                
One of the main problems in assigning a richness to a galaxy 
cluster in the optical
is the fact that the cluster is seen against a background galaxy distribution    
which is far from being homogeneous but shows structure on all scales.          
The latter effect is clearly shown by the autocorrelation analysis              
of the two-dimensional projected galaxy distribution on the sky. It        
is therefore difficult to determine a background subtracted galaxy 
number of a cluster in a unique fashion.      
Also the so called projection effects, in which several galaxy groups or a      
filamentary structure in the line of sight can mimic a compact rich cluster,    
are basically a result of this inhomogeneous matter distribution
(e.g. van Haarlem 1997).           
                                                                                
The possibility to detect galaxy clusters in X-rays has since been recognized   
as a way to improve the unambiguity of the detection. 
The X-ray emission observed in clusters         
originates from the thermal emission of hot intracluster gas (e.g. Sarazin      
1986) which is distributed smoothly throughout the cluster. The plasma is       
bound by the gravitational potential well of the clusters and fills             
the potential approximately in a hydrostatic fashion. Therefore the plasma      
emission is a very good tracer of the cluster's gravitational potential.         
Even though the plasma is very tenuous, the large volume makes galaxy clusters      
the most luminous X-ray sources besides AGN. In addition the thermal            
emission for the typical intracluster plasma                                    
temperatures of several keV has the                                             
spectral maximum in the soft X-ray band where the available X-ray              
telescopes are most effective. This makes galaxy clusters readily               
detectable out to large distances with present X-ray telescopes.                
                                                                                
However, the main advantage of the X-ray detection is the fact that
the X-ray luminosity is closely correlated to the cluster mass
(Reiprich \& B\"ohringer 1999), with a dispersion of about 50\% 
in the determination of the mass for a given X-ray luminosity
(Reiprich \& B\"ohringer, in preparation). Thus, in summary X-ray 
selection provides the following positive features:

\noindent                                                                       
{\bf $\circ$ } An effective selection
by mass (with a known dispersion which can be taken into account in any 
corresponding modeling). 
                                                                                
\noindent                                                                       
{\bf $\circ$ } The X-ray background originates mostly from distant point        
sources which are very homogeneously distributed (e.g. Soltan \& Hasinger 1994). 
Therefore the X-ray background is very much easier to subtract from the cluster
emission than the optical galaxy background distribution.

\noindent                                                                       
{\bf $\circ$ } The X-ray surface brightness is much more concentrated towards 
the cluster centre as compared to the galaxy distribution. 
Therefore the effect of overlaps along the line of sight is minimized.
                                                                                
The construction of statistically complete samples of X-ray clusters 
started with the completion of the first all-sky          
X-ray survey by the HEAO-1 satellite (Piccinotti et al.\ 1982, Kowalski et      
al.\ 1984). With additional observations from EINSTEIN and EXOSAT               
a cluster sample of the $\sim 50$ X-ray brightest objects with more     
detailed X-ray data was compiled (Lahav et al.\ 1989, Edge et al.\ 1990)   
and with the analysis of deeper EINSTEIN observations            
the first deep X-ray cluster survey, within the EMSS, has been obtained
(Gioia et al. 1990, Henry et al. 1992). The latter     
survey allowed in particular to address the question of the evolution of
cluster abundance with redshift (e.g. Henry et al.\ 1992, Nichol et al. 1997).  
The ROSAT All-Sky Survey (RASS), the first X-ray all-sky survey conducted
with an X-ray telescope (Tr\"umper 1992, 1993) provides an ideal basis
for the construction of a large X-ray cluster sample for cosmological studies.  
Previous cluster surveys based on the RASS include: Allen et al.\ (1992),
Romer et al.\ (1994), Pierre et al. (1994), Crawford et al.\ (1995, 1999), 
Burns et al.\ (1996), Ebeling et al.\ (1996, 1998, 2000), 
De Grandi (1999a,b), Henry et al.\ (1997),
Ledlow et al.\ (1999), B\"ohringer et al. (2000a), and Cruddace et al. (2000).
                                                                                
The goal of the present survey work is to fully exploit the 
RASS for the search of     
clusters as far as one can hope to obtain a fairly complete cluster sample       
and a reasonably good characterization of the X-ray source properties.
For the construction of such a cluster sample optical follow-up    
observations, in addition to the X-ray analysis and X-ray/optical correlations,   
are necessary to clearly identify the nature of the X-ray sources
and to determine the cluster redshifts.        
To this aim we have conducted an intensive follow-up optical survey project
as an ESO key program from 1992 to 1999
(e.g. B\"ohringer 1994, Guzzo et al. 1995, B\"ohringer et al. 1998,
Guzzo et al. 1999) which has been termed REFLEX 
(ROSAT-ESO-Flux-Limited-X-ray) Cluster Survey.
Within this program the identification of all the cluster candidates
at $\delta \le 2.5\deg $ and
down to a flux limit of $3 \cdot 10^{-12}$ erg s$^{-1}$ cm$^{-2}$ 
in the ROSAT band (0.1 to 2.4 keV) has been completed. This
sample includes 452 identified galaxy clusters, 449 of which have a measured
redshift. An extension of the identification programme 
down to a lower flux limit is planned and a large number of redshifts
for this extension has already been secured.

A complementary RASS cluster
redshift survey programme is conducted for the northern celestial hemisphere
in a collaboration by MPE, STScI, CfA, and ESO, the Northern ROSAT All-Sky
Cluster Survey (NORAS; e.g. B\"ohringer 1994, Burg et al. 1994) 
and a first catalogue containing 483 identified X-ray galaxy clusters
has recently been published (B\"ohringer et al.\ 2000a).
It is the future aim to combine the northern and southern surveys
which at present are based on slightly different identification
strategies, mostly due to the different optical data available.
We have also successfully extended the cluster search into
the region close to the galactic plane covering about 2/3 of the region
with galactic latitude $|b_{II}| < 20\deg$ (B\"ohringer et al. 2000c). 
                                                                                
In this paper we describe the selection of the cluster candidate sample         
for the REFLEX Survey. Earlier results coming from a subsample of a 
preliminary RASS I based version of the REFLEX cluster catalogue 
comprising 130 clusters at a flux limit of 
$3 -4 \cdot 10^{-12}$ erg s$^{-1}$ cm$^{-2}$ (as measured in the 
0.5 - 2 keV energy band) in 2.5 sr of the southern sky 
have been published by De Grandi et al. (1999a,b).
The layout of the paper is as follows. In section 2 we characterize              
the depth and the sky area of the study and in section 3 the basic RASS data    
used as input. It was found that a reanalysis of the X-ray properties of the    
clusters in the RASS was necessary for the project. This new reanalysis         
technique and its results are presented in section 4.              
The selection of the galaxy cluster candidates by means of a correlation
of the X-ray source positions with the optical     
data base from COSMOS is described in section 5 and 6. The further X-ray source 
classification is discussed in section 7. Section 8 provides tests of the
sample completeness. The resulting REFLEX cluster sample for a flux limit 
of $3 \cdot 10^{-12}$ erg s$^{-1}$ cm$^{-2}$ and some of its characteristics
is described in section 9. Further statistics of the X-ray properties
of the REFLEX clusters and the contamination of the sample by non-cluster
sources is discussed in section 10 and section 11 provides a summary 
and conclusions.   
             
\section{The REFLEX galaxy cluster survey}                                
                                                                                
The survey area of REFLEX is the southern hemisphere below a declination     
of +2.5$\deg $. A region of $\pm 20$ degrees around the galactic plane is       
excluded form the study, since clusters are difficult to recognize              
optically in the dense stellar fields of the Milky Way and the X-ray detection  
is hampered by the higher interstellar absorption in the inner parts              
of the galactic band.                                                           
The region 2.5 degrees above the equatorial                            
equator is included in this project since the COSMOS data extend         
up to this declination. It provides some overlap with     
the NORAS survey project (e.g. B\"ohringer et al. 2000a) 
where both cluster
identification programmes can be compared. The total area thus covered          
is 4.34 ster or 14248 deg$^{2}$.                                                
                                                                                
In addition to the region around the galactic plane, the dense stellar      
fields of the two Magellanic clouds prevent an efficient galaxy search          
in these regions of the southern sky. In particular the star-galaxy             
separation technique used in the construction of the COSMOS                     
data base became inefficient in these crowded areas (H.T. MacGillivray,         
private communication) and therefore no star-galaxy classification              
is actually provided in the COSMOS data released and used for our               
project.                                                                        
Therefore we exclude an area                                                    
of 244.4 deg$^{2}$ for the LMC and 79.8 deg$^{2}$ for the SMC which             
essentially follows the boundaries of those UK-Schmidt plates without           
object classification.                                                          
The areas which are excised from our survey are specified in detail in Table 1. 
                                                                                
   \begin{table}                                                                
      \caption{Regions of the sky at the LMC and SMC excised from the Survey}   
         \label{Tempx}                                                          
      \[                                                                        
         \begin{array}{llll}                                                    
            \hline                                                              
            \noalign{\smallskip}                                                
 {\rm region}& {\rm RA range }  & {\rm DEC range} & {\rm area (ster)} \\        
            \noalign{\smallskip}                                                
            \hline                                                              
            \noalign{\smallskip}                                                
{\rm LMC 1}  & 58  \to 103^o  & -63 \to -77^o   & 0.0655 \\                     
{\rm LMC 2}  & 81  \to 89^o  & -58 \to -63^o   & 0.0060 \\                      
{\rm LMC 3}  & 103 \to 108^o  & -68 \to -74^o   & 0.0030 \\                     
{\rm SMC 1}  &358.5 \to 20^o  & -67.5 \to -77^o & 0.0189 \\                     
{\rm SMC 2}  &356.5 \to 358.5^o & -73 \to -77^o   & 0.0006 \\                   
{\rm SMC 3}  & 20  \to 30^o   & -67.5 \to -72^o & 0.0047 \\                     
            \noalign{\smallskip}                                                
            \hline                                                              
         \end{array}                                                            
      \]                                                                        
   \end{table}                                                                  
%
%
                                                                                

The total survey area after                                                     
this excision amounts to 4.24 ster or 13924 deg$^{2}$ which             
corresponds to 33.75 \% of the sky.                         
This survey covers the largest area for which currently a                       
homogeneous combined optical/X-ray survey is possible,
since there is no optical survey                        
covering both hemispheres simultaneously. 
                                                      
The observational goal of this survey programme is the identification       
and redshift determination of all galaxy clusters in the study area
above a given flux limit. In a first step, within the ESO key programme,
we have completed the observations for a sample of 452 galaxy
cluster (with redshifts for 449 clusters) above a limiting flux of 
$3 \cdot 10^{-12}$ erg s$^{-1}$ cm$^{-2}$. In addition we have already 
secured many redshifts at lower fluxes and we plan to extend the 
redshift survey
to flux limit of $1.6 - 2 \cdot 10^{-12}$ erg s$^{-1}$ cm$^{-2}$. This
corresponds to a count rate limit in the hard ROSAT band of about 
$0.08 - 0.1$ cts  s$^{-1}$. With a typical exposure in the southern
part of the RASS of about 330 sec this yields about $25 - 30$ photons for the 
fainter sources. This is still just enough to determine a flux within
uncertainty limits of typically less than 30\% and provides some leverage for
the determination of some source properties. At this flux limit we expect 
between 700 and 1000 galaxy clusters in the survey area (based on
the number counts of previous surveys e.g. Gioia et al. 1990,
Rosati et al. 1998).

For the preparation of the candidate sample we have therefore chosen
to start with a source sample with a count rate limit of 0.08 cts  s$^{-1}$ 
in the hard ROSAT band (channel 52 to 201 corresponding approximately to
an energy range of 0.5 to 2.0 keV). Note that all the fluxes quoted in
this paper refer to the total ROSAT energy band (0.1 - 2.4 keV) in
contrast to the more restricted band of pulse high channels chosen 
for the determination of the count rate.
This count rate limit translates into 
a flux limit for cluster type spectra of $1.55 - 1.95 \cdot              
10^{-12}$ erg s$^{-1}$  cm$^{-2}$ a range determined mainly by
variations of the interstellar column density in the REFLEX area 
(20\% in the range $1 - 10 \cdot 10^{20}$ cm$^{-2}$). 
Weaker dependences on the cluster temperature (e.g. $\sim$ 1.4\%                 
in the range 3 - 8 keV, see Fig. 8 in B\"ohringer et al. 2000a)
and redshift (in analogy to the optical K-correction,   
0.5\% in the range z = 0 to z = 0.2) are found. 
(Below about 2 keV the temperature
dependence is stronger, however). We will be quoting 
unabsorbed flux values in         
the ROSAT energy band (defined as 0.1 to 2.4 keV) throughout this paper
since the results in this energy band are less dependent on the 
spectral model assumptions for the sources compared to any other 
significantly wider band definition.  
Further assumptions or information on the source spectrum (e.g. intracluster   
plasma temperatures) are                                                    
needed to subsequently convert these primary data to other energy               
bands or to bolometric fluxes and luminosities.                                

For the calculations of the fluxes, the luminosities, and some other physical
parameters in this paper we have made the following assumptions. A first
approximate unabsorbed flux is calculated for each 
X-ray source from the observed count rate,
prior to any knowledge about its nature and redshift by assuming a thermal 
spectrum with a temperature of 5 keV, a metallicity of 0.3 solar 
(with abundances taken from Anders and Grevesse (1989). A redshift of 0,
and an interstellar column density of hydrogen as obtained from Dickey
\& Lockman (1990) and Stark et al. (1992) for the X-ray source position
is adopted. 
This nominal flux is used to impose the flux limit on the X-ray source sample.
After a cluster has been identified and its redshift secured
a better temperature estimate is obtained by means of the 
temperature/X-ray luminosity relation (Markevitch 1998)
\footnote{We are using the temperature/X-ray luminosity relation
uncorrected for the effects of cooling flows, since the REFLEX X-ray
data used here are also not corrected for the possible effects of
cooling flows.}, 
and a corrected flux
and X-ray luminosity is calculated taking the new estimated temperature,
the K-correction for the observed redshift, and the dependence on the 
interstellar absorption into account. The X-ray luminosities are always
calculated in the ROSAT band in the cluster restframe, while the fluxes
are given in the ROSAT band for the observer frame as unabsorbed fluxes.
The calculations are performed within the EXSAS software system 
(Zimmermann et al. 1994) with the spectral code from John Raymond
(Raymond \& Smith 1977). Instead of using the standard codes of EXSAS
for the count rate flux conversion we are using our own macros which
have been tested against XSPEC and show a general agreement within less 
than 3\%. For the calculations of the luminosities and
other physical properties of the clusters we assume a standard cosmology 
with $H_0 = 50$ km s$^{-1}$ Mpc$^{-1}$, $\Omega_0 = 1$ and $\Lambda = 0$.
While the basis of the source detections is the standard analysis
of the RASS (Voges et al. 1999) we have reanalysed the source count rates
and other source properties as described in section 4 with the growth
curve analysis technique. Note that previous
comparisons of the results of this technique with deeper pointed
ROSAT observations show that the measured flux underestimates the total
cluster flux typically by an amount of 7-10\%
(B\"ohringer et al. 2000a). 
The fluxes and luminosities quoted here are the measured values 
without a correction for the possibly missing flux. 

                                                                                
\section{The ROSAT All-Sky Survey data}                                         
                                                                                
The starting point of the sample construction is the RASS 
standard analysis source list.
This source list was constructed during the second RASS                         
processing (RASS II) by Voges et al.\ (1996, 1999)
using subsequently the LDETECT, MDETECT, and                                    
Maximum-likelihood algorithms (referred to as Standard Analysis Software        
System).                                                             
While only highly significant sources                                           
(maximum likelihood parameter  $L \ge 15$) with                                 
count rates above 0.05 cts s$^{-1}$ and with interactively             
confirmed existence  entered into the published RASS bright                     
source catalogue (Voges et al.\ 1999), the primary, ROSAT Team internal 
source catalogue down to a                                         
source likelihood of $L = 7$ is used here. 
At this low likelihood the significance                                         
of some of the sources is below $3\sigma $ and not all of the sources are       
expected to be real. However, this low threshold guarantees that no   
sources are missed in the final sample, after the new flux cut          
is introduced.                                                                  
In total 54076 sources were found by the standard analysis in RASS II
down to a likelihood of 7 in the study area of REFLEX (see Table 3).

\begin{figure}                                                                  
\psfig{figure=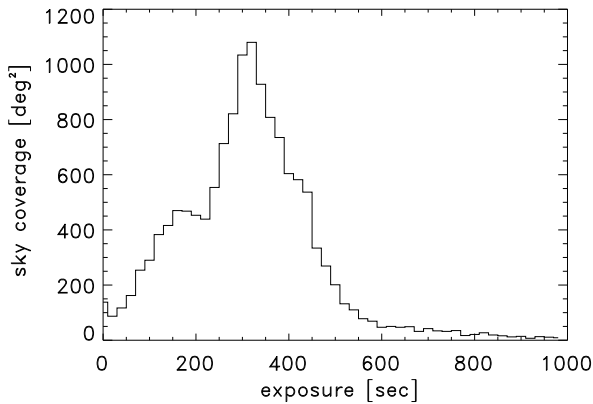,height=6.0cm}                                         
\caption{Exposure time distribution of the ROSAT All-Sky Survey as analyzed     
in RASS II in the  area of the REFLEX survey}\label{fig1}
\end{figure}

One complication in using RASS data is the non-homogeneous sky coverage.           
Since the sky was scanned in great circles perpendicular
to the ecilptic,        
the shortest exposures are near the ecliptic equator, while this piles     
up at the ecliptic poles. In addition the satellite had to be switched-off      
frequently during the passages through the radiation belts. This affects  
in particular the southern sky data, since due to the South Atlantic Anomaly     
of the Earth magnetic field the radiation belts are more prominent       
in the southern sky at the flight altitude of ROSAT. Some minor regions are         
underexposed because the data have been rejected for reasons of bad quality   
of the attitude solution. This leaves low exposure areas in the                 
primary data. The  resultant exposure distribution is shown in 
Fig.~\ref{fig1}. The mean exposure is 335 s and the median         
is 323 s (compared to NORAS with mean and median exposures of
397 and 402 s, respectively). Table 2 gives the  
fractions of the sky area which                 
are underexposed. Only a few percent of the sky area are strongly underexposed  
and only about 12\% has less than half the median exposure.       
Such structure imposed by the exposure drop-outs is therefore not so dramatic,
but has to be taken into account for any statistical analysis of the cluster
population. Its actual effects depend on the X-ray flux limit of the sample.

   \begin{table}                                                                
      \caption{Fractions of the REFLEX survey area with low exposure}   
         \label{Tempx}                                                          
      \[                                                                        
         \begin{array}{lr}                                                      
            \hline                                                              
            \noalign{\smallskip}                                                
 {\rm exposure~~ [s]}& {\rm fraction~ of~ the~ sky~ area }  \\                  
            \noalign{\smallskip}                                                
            \hline                                                              
            \noalign{\smallskip}                                                
 <50   & 0.019 \\                                                               
 <100  & 0.054 \\                                                               
 <150  & 0.117  \\                                                              
 <200  & 0.200 \\                                                               
 <300  & 0.413 \\                                                               
            \noalign{\smallskip}                                                
            \hline                                                              
         \end{array}                                                            
      \]                                                                        
   \end{table}                                                                  
%
%
                                                                                
\begin{figure*}                                                                 
                                                                                
\centerline{\psfig{figure=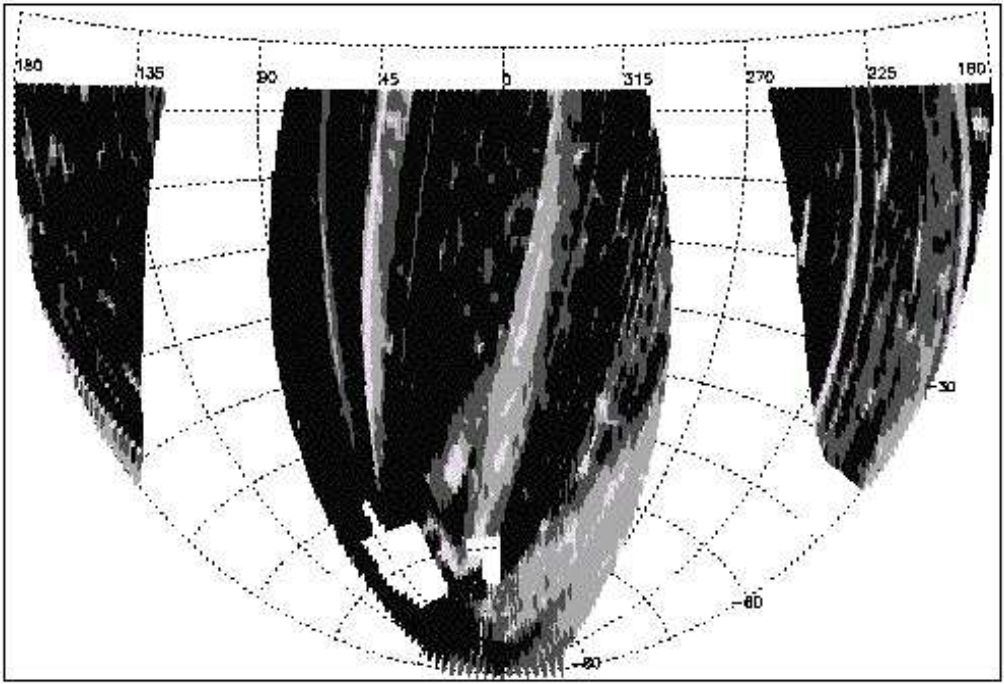,height=9.5cm}}                             
\caption{Exposure distribution in the area of the REFLEX survey. Four grey levels
have been used for the coding of the exposure times, with increasing intensity      
for $t_{exp} < 100$ s, $100 < t_{exp} < 200$ s , $200 < t_{exp} < 300$ s, 
$t_{exp} \ge 300$ 
respectively.}\label{fig2}
\end{figure*}

In particular for the clustering analysis the  
distribution of underexposed areas has to be known, so that it    
can be taken into account. This distribution is shown in Fig.~\ref{fig2}.
The underexposed area is not contiguous, but it is more or       
less confined to four strips in the southern sky. These strips reflect          
the shut-off times of the ROSAT-detector during the passage of the radiation
belts in the South Atlantic Anomaly.

   \begin{table}                                                                
      \caption{Number of X-ray sources obtained in the subsequent selection steps:
      regional selection, count rate cut, removal of multiple detections
      (in two steps), removal of bright stars (prior to the correlation
       with the COSMOS data base), final flux cut. This selection does not yet
       involve the search for cluster candidates which is described in section 5.}
         \label{Tempx}                                                          
      \[                                                                        
         \begin{array}{lr}                                                      
            \hline                                                              
            \noalign{\smallskip}                                                
 {\rm selection} & {\rm number~ of~ sources }\\                                   
            \noalign{\smallskip}                                                
            \hline                                                              
            \noalign{\smallskip}                                                
{\rm RASS II~ (L \ge 7)~ in~ study~ area }  &  54076 \\                         
{\rm count~ rate~ cut \ge 0.08 cts~ s^{-1} }      &   6593   \\                   
{\rm removal~ of~ double~ detections~}  & \\
{\rm \hskip 1cm (d~ \le~ 2 arcmin)^{\dagger} }  &  4410   \\                     
{\rm removal~ of~ multiple~ cluster~} & \\
{\rm \hskip 1cm detections~ (d~ >~ 2 arcmin)  }  &  4206  \\            
{\rm removal~ of~ bright~ stars  }  &  3754  \\
{\rm flux~ limit~~ 3\cdot 10^{-12} erg~ s^{-1} cm^{-2} } & 1169\\       
            \noalign{\smallskip}                                                
            \hline                                                              
         \end{array}                                                            
      \]                                                                        
\begin{list}{}{}                                                               
\item[$^{\dagger}$]  d is the positional separation of multiple detections                                   
\end{list}                                                                     
   \end{table}                                                                  

\section{Reanalysis of the X-ray data of the RASS sources with                  
the GCA technique}

Since we found from our previous studies that the flux of               
extended sources is underestimated by the standard RASS source detection        
algorithm (Ebeling et al.\ 1996, De Grandi et al.\ 1997),                 
a reanalysis of the source fluxes is necessary before            
we introduce a count rate cut or flux limit in the source list, that 
will subsequently serve   
as the basis for the construction of the X-ray flux-limited cluster sample.
This is especially important in the present study since the majority
of the cluster sources feature a significant extent and many appear
not perfectly spherically symmetric.
To this end we have developed a new source characterization technique,
the growth curve analysis (GCA) method, 
which is described in detail in B\"ohringer et al. (2000a).
The strategy for the development                  
of the new algorithm is to obtain reliable fluxes 
for extended sources and to extract as much useful information                 
from the raw data as possible with a simple and                    
robust technique. The simplicity of the technique is particularly             
important in devising a model for the source detection               
from theoretically constructed catalogue data in order to                      
simulate possible selection effects in the sample.                
We have given preference to use the GCA method for this analysis
over the methods described by Ebeling et al.\ (1996) and 
De Grandi et al.\ (1997) since it makes more extensive use of the 
photon count information, is not based on the assumption of a particular cluster
model and therefore free of this type of bias, and provides reliable count
rates also for lower fluxes. A more detailed comparison will be given elsewhere.
Here we give only a brief outline on the GCA.

For each source GCA returns (among other                         
information) the following most important parameters which will be               
used in the source selection work:                                              
                                                                                
{\bf $\circ$} {observed source count rate (background subtracted)}              
                                                                                
{\bf $\circ$} {Poisson error (photon statistics) for the count rate}                
                                                                                
{\bf $\circ$} {locally redetermined center of the source}                       
                                                                                
{\bf $\circ$} {mean exposure for the source region}                             
                                                                                
{\bf $\circ$} {significance of the source detection}                            
                                                                                
{\bf $\circ$} {estimate of the radius out to which the source emission              
is significantly detected}                                                      
                                                                                
{\bf $\circ$} {extrapolated source count rate (obtained by model fitting to     
the source emission distribution)}                                              
                                                                                
{\bf $\circ$} {hardness ratio characterizing the source spectrum and            
its photon statistical error}                                                   
                                                                                
{\bf $\circ$} {fitted source core radius}                                       
                                                                                
{\bf $\circ$} {Kolmogorov-Smirnov test probability that the source shape 
is consistent with a point source}           
                                           
In the following we give a very brief summary of the GCA method.
The basic parameters are derived for the photon distribution                    
in the three energy bands ``hard'' (0.5 to 2.0 keV, channels 52 - 201),         
``broad'' (0.1 to 2.4 keV, channels 11 - 240), and ``soft'' (0.1 to 0.4          
keV, channel 11 - 40). The band definitions are the same as those used
in the standard analysis (Voges et al. 1999).
Here we are only using then hard band results, since the clusters 
are detected in this band with the highest signal to noise ratio. 
An exception is the hardness ratio which 
requires the results from the hard and soft bands. 
                                                                                
\begin{figure}                                                                  
\psfig{figure=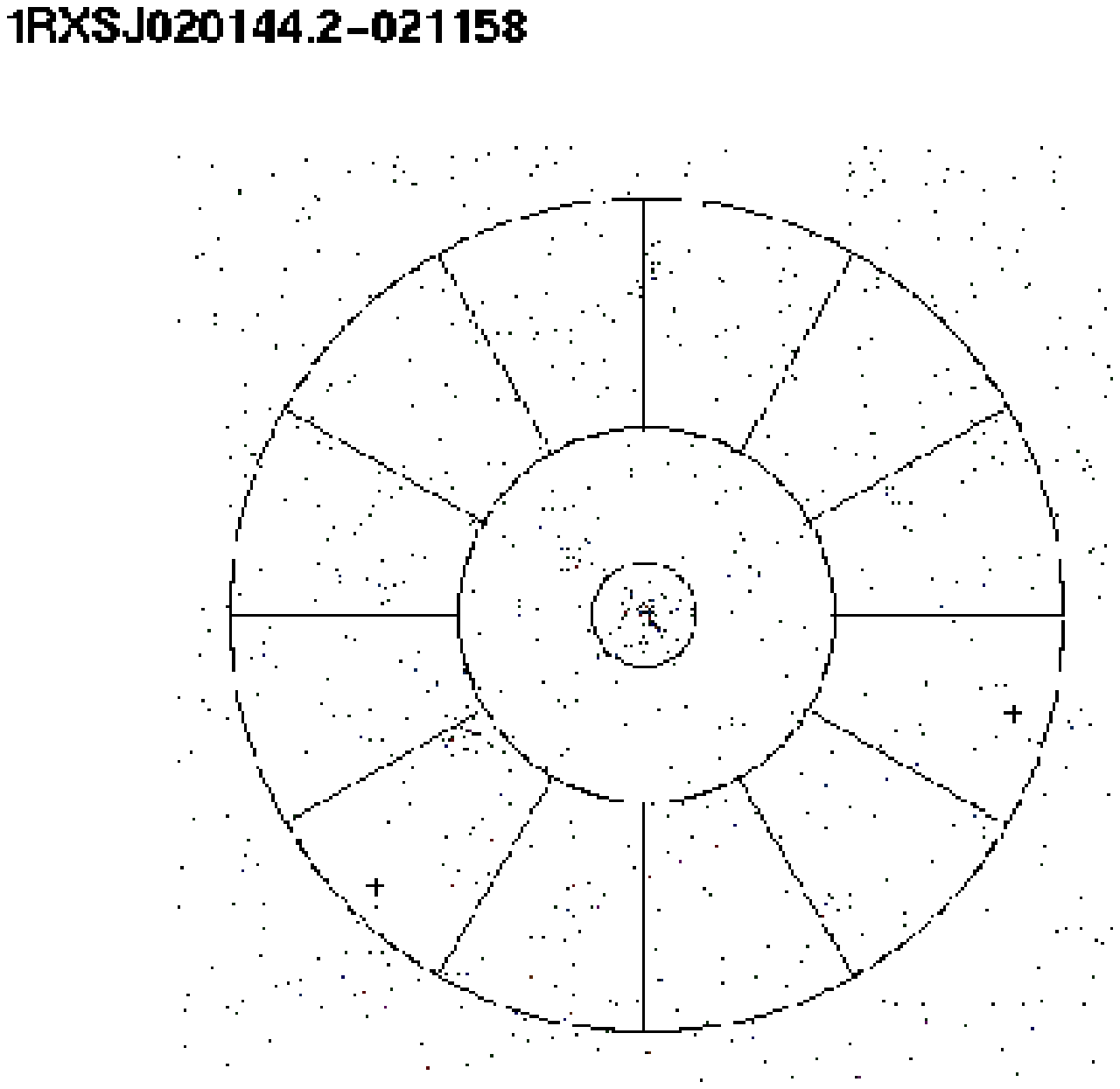,height=7.0cm}                                           
\caption{Example of the set-up of the source characterization method used      
in the GCA technique. The image shows the hard band photon distribution from
an area of the RASS in a 1.5 degree box around the X-ray source.
The outer two circles enclose the area of the background             
determination. This background area is divided into 12 sectors. The             
two sectors marked by a cross are discarded from the background                 
determination. They are flagged by a $2.3\sigma$ clipping technique
indicating a possible contamination or strong fluctuation (see
B\"ohringer et al. 2000a for details). The inner ring marks the outer          
radius out to which significant X-ray emission from the source is detected.      
}\label{fig3} 
\end{figure}                                                                    
                                                                                
\begin{figure}                                                                  
\psfig{figure=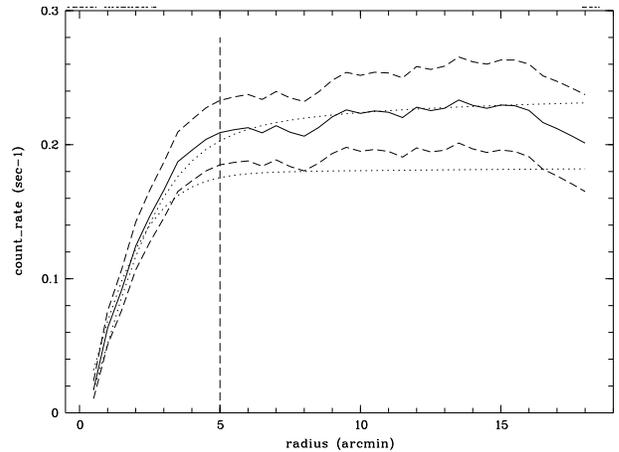,height=6.0cm}                                      
\caption{Integrated count rate profile for the source shown in Fig. 3.           
The integrated count rate profile is background subtracted and the two          
dashed curves give the photon statistical error ($1 \sigma$)                    
of the count rate which                                                         
includes the uncertainty of the signal and background determination.            
The vertical dashed line shows the outer source radius as explained in the      
text. The lower dotted line shows the $\chi ^2$ fit of a point source           
to the data while the upper dotted line shows the best King model fit.}\label{fig4}
\end{figure}

The source count rate is determined from the
cumulative, radial source count rate profile (''growth curve'') 
after background subtraction. The construction of the growth curve
is preceded by a redetermination of the source center and by
the background measurement. 
As a typical example, the growth
curve for the source shown in Fig.~\ref{fig3} is displayed in Fig.~\ref{fig4}. 
In addition to the count rate as a function of 
integration radius, the uncertainty limits determined by
photon statistical error (including the error for the background 
subtraction) also are calculated and displayed in Fig.~\ref{fig4} as 
dashed lines.

The count rate is determined in two alternative ways. In the first 
determination an outer radius of significant X-ray emission, $r_{out}$,
is determined
from the point where the increase in the $1\sigma$ error is larger
than the increase of the source signal. The integrated count rate
is then taken at this radius. In the second method a horizontal
level is fitted to the outer region of the growth curve (at $r \ge r_{out}$),  
and this plateau is adopted as the source flux. We use the second
approach as the standard method but use also the first method as a check,
and a way to estimate systematic uncertainties in the count rate
determination in addition to the pure photon statistical errors.
We also determine a fitted total count rate by means of a $\beta$-model
as described below.
For sources where close neighbours disturb the count rate
measurement we run a separate deblending analysis.

The two most important source quality parameters determined within GCA are the
spectral hardness ratio and the source extent.
The hardness ratio, $HR$, is defined as
$HR = {H - S \over H + S}$                                                     
where $H$ is the hard band and $S$ the soft band source count rate
(both determined for the same outer radius limit).

The source extent is investigated in two ways.
In the first analysis a $\beta$-model profile (Cavaliere \& Fusco-Femiano 1976)
convolved with the averaged survey PSF 
(G. Hasinger, private communication) is fitted to the differential
count rate profile (using a fixed value of $\beta$ of $2/3$ ) yielding  
a core radius, $r_c$, and a fitted total count rate. Secondly, a
Kolmogorov-Smirnov test is used to estimate the probability that
the source is consistent with a point source . The source is flagged to
be extended when the KS probability is less than
0.01. Tests with X-ray sources which have been identified with
stars or AGN show a false classification rate as extended sources
of about 5\% (these results will be discussed in detail in a 
subsequent paper).

All 54076 RASS II sources in the REFLEX study region were subjected to the
GCA reanalysis. All sources with a count rate $\ge 0.08$ cts s$^{-1}$
were retained for the primary sample.

\begin{figure}                                                                  
\psfig{figure=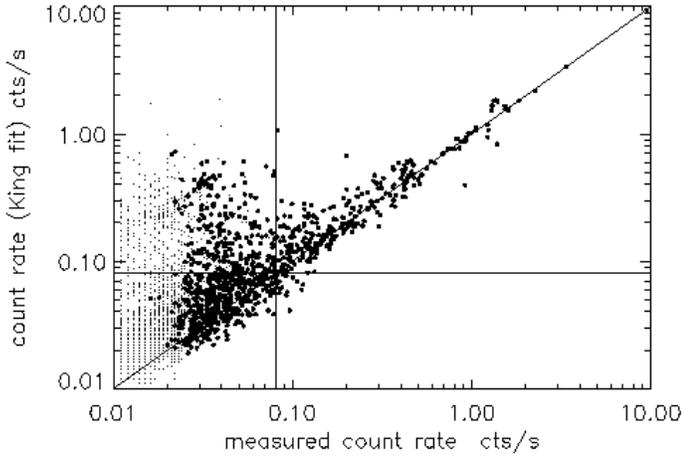,height=6.5cm}                                        
\caption{Comparison of the count rates determined for the RASS sources          
in the REFLEX study area with both techniques, the ``measured''                 
count rate out to the radius of significant X-ray emission and                  
the count rate obtained by fitting a King profile to the source                 
count rate profile shape. The diagonal line gives the location of the points    
for which both measures are equal. The vertical and horizontal line give        
the count rate cut values for the two techniques, respectively. The data        
for which the significance of the detected signal is found to be greater than   
3 are marked by heavy dots, while the data below this significance threshold    
are plotted by light dots. In the graph for clarity only the first 5000        
sources of the total sample of 54076 sources are shown.}\label{fig5}                        
\end{figure}

For the first sample cut in count rate we have been very conservative.
In addition to selecting all sources with a count rate  
$\ge 0.08$ cts s$^{-1}$ as measured at $r_{out}$ we have also retained
all sources featuring a fitted total source count rate above this
value in the $\beta$-model fit  and a
significance for the source detection $\ge 3\sigma$. While this leads
to  the inclusion of a significant fraction of sources below the count
rate cut (due to less successful  $\beta$-model fits) it also ensures
that sources with pathological count rate profiles featuring an
underestimate of  $r_{out}$ are not lost before all sources can individually
be inspected in the GCA diagnostic plots.   
A comparison of the GCA determined
count rate (first method) and the fitted count rate
is shown in Fig.~\ref{fig5}. 
There is a good correlation of the two count rate values above a measured count 
rate of about 0.1 cts s$^{-1}$. At low values of     
the GCA count rate, the fitted count rates show a large scatter. This 
is mostly due to the poor photon statistics providing not enough
constraints on the source shape for a good enough $\beta$-model fit.
A closer inspection of the results shows that at low count rates the
fitted results give overestimates in more than one fourth of the sources,  
leading to an 
oversampling of about 20\%. This is of no harm to
the final sample construction, since the final REFLEX sample is
obtained by another cut in flux well above this limit.

In total the first count rate cut leads to a sample of 6593 sources.
This sample contains still a large number of original multiple detections
of extended sources by the RASS II standard analysis. The new 
analysis method offers an efficient way
of removing most of these multiple detections. 
In the redetermination of the source center in the GCA analysis,           
the technique usually finds a common center for the multiple detection
of clusters within small numerical  
differences in the position (generally $< 1$ arcmin). 
Since at a separation of 
2 arcmin also point sources are already overlapping, we have used
a maximal separation of 2 arcmin to identify these multiple detections.
Removing the redundant detections 
the source list shrinks to 4410 sources. 
This is the sample that was subjected to the first X-ray optical 
correlation as described in the next section.
Further screening revealed another 204 redundant detections in very extended
clusters where the method has settled in different local maxima
(with a separation larger than 2 arcmin), but
which are easily recognized visually (see Table 3 which summarizes the 
number of sources obtained in the subsequent 
steps of the sample construction).

\begin{figure}
\psfig{figure=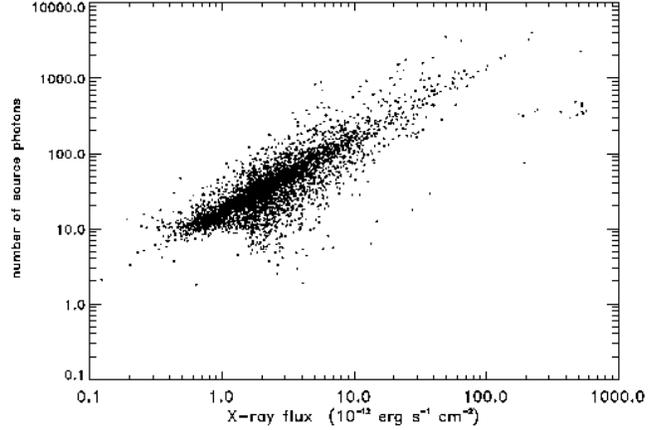,height=6.0cm}                                            
\caption{Distribution of the number of source photons (background subtracted)
obtained as a function of the X-ray flux}\label{fig6} 
\end{figure}

To illustrate how well the X-ray properties can still be characterized
near the flux limit of this sample we provide some statistics on
the source photon number and detection significance for the sources in 
the sample. This is interesting in the light of the discussion in section
2, where we outline the strategy for the survey depth and where we argue
that the depth of a survey is limited, if we require a certain accuracy
for the derived X-ray properties requiring a minimum
source photon number. In Fig.~\ref{fig6} we show the number of 
photons detected for the
sample of 4206 RASS II sources above the count rate cut. 
Most sources at the count rate limit have more than 10 photons
allowing for a flux determination with an accuracy 
of at least 30\%.  Note
that besides the ''main sequence'' of data points there is also
a fraction of sources well below these typical data. These data points
come from the low exposure areas in RASS II. To avoid unwanted selection
effects in the sample it is useful to introduce a source count limit,
and to correct for this cut in the sample selection function
as discussed later.

\begin{figure}
\psfig{figure=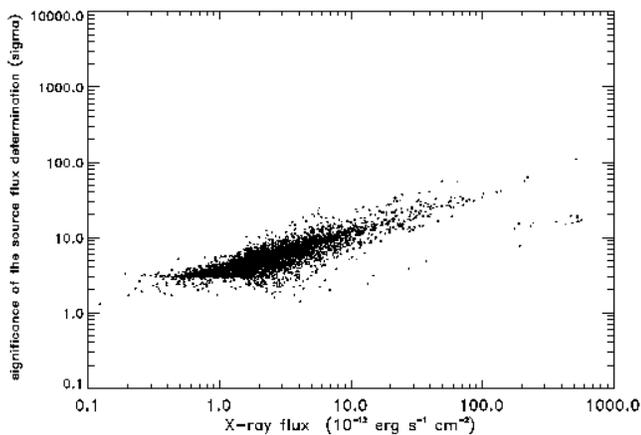,height=6.0cm} 
\caption{Distribution of the significance of the flux determination 
as a function of the X-ray flux. For the definition of the significance 
parameter see the text. 
}\label{fig7}
\end{figure}

Fig.~\ref{fig7} shows in a similar way the typical significance for the flux
measurement of a source as a function of the X-ray flux. This significance 
parameter is defined as $S/N = N_s / \sqrt{N_s + N_b}$, where $N_s$ are the
source counts and $N_b$ are the background counts in the source region. 
Again there are some sources with a low 
significance for the source flux determination, which come from the 
low exposure areas. 

\section{Correlation of the X-ray source list with the COSMOS data base}        
                                                                                
\subsection{The Cosmos data base}

Since the X-ray properties which are described above do not allow by 
themselves an identification of the X-ray sources associated to clusters,
we have to include
information from an optical data base in the further identification 
process. For this we are using the most comprehensive optical data base
covering the southern sky and the area of the REFLEX survey: the COSMOS
scans of the UK-Schmidt survey plates (MacGillivray \& Stobie 1984). 
There are also the complementary APM scans of the same photographic
survey material, but the galaxy classification in the APM survey
concentrates on the southern part of the sky south of the galactic plane
(Maddox et al. 1990) which covers only about 2/3 of the REFLEX region.

The UK-Schmidt survey has been performed using IIIa-J photographic
plates at the 1.2m UK-Schmidt-telescope. 
The plates were scanned 
within a sky area of about $5.35\deg \times 5.35\deg$ per plate
with the fast COSMOS scanning machine and subsequently analysed 
yielding 32 parameters for the source characterization
per object. These parameters describe the object position,
intensity, shape, and classify the type of object.  
Object images are recognized down to about $b_J \sim 22$ mag.
This allows a subsequent star/galaxy separation which
has been estimated to be about 95\% complete with about 5\% contamination
to $b_J \sim 19.5$ mag and about 90\% complete with about 10\%
contamination to  $b_J \sim 20.5$ mag (Heydon-Dumbleton, Collins,
\& MacGillivray 1989, Yentis et al. 1992, MacGillivray et al. 1994,
and Mac Gillivray priv. communication). The galaxy magnitudes were
intercalibrated between the different plates using the substantial
plate overlaps and absolutely calibrated by CCD sequences 
(Heydon-Dumbleton et al. 1989, MacGillivray et al. 1994).    
                                 
\subsection{Correlation of the X-ray sources with the COSMOS galaxy distribution}
                                               
In search of cluster candidates as counterparts to the RASS sources we          
correlate the X-ray source positions with the galaxy catalogue of
the COSMOS data base. The             
basis of this correlation are counts of galaxies in circles around the            
X-ray source positions to search for galaxy density enhancements.               

Here we should make some remarks about the strategy behind the choice of 
the present cluster search algorithm. As mentioned before 
it is difficult to devise a good algorithm to select
the most massive clusters of galaxies from optical sky survey images. 
We use a comparatively simple algorithm (aperture counts as 
compared to e.g. matched filter techniques). This simple technique 
seems well adapted to our needs and the depth of the COSMOS data set: 
(i) the technique is used to only flag the candidates and
there is no need to determine a cluster richness, since we use
the X-ray emission for a quantitative measure of the clusters;
(ii) while matched filter techniques may introduce a bias, since 
{\it a priori} assumptions are made about the shape of an idealized,
azimuthally symmetric cluster, we are interested in introducing
as little bias and as few presumptions as possible; (iii) the actual numbers
in the galaxy counts are limited and therefore the shape matching is not
precise and is affected by low number statistical noise. Therefore
our technique is not seen as a perfect and objective
cluster characterization algorithm. The cluster selection should primarily 
depend on the X-ray criteria. We have chosen a very low cut
for the optical selection which results in a substantially larger
candidate sample compared to the expected number of clusters,
with an estimated contamination of as much as 30 - 40\%.
But it assures
on the other hand that we have a highly complete candidate sample.
This overabundance of candidates is thus a necessary condition to 
obtain an essentially X-ray selected sample for our survey. 

The galaxy counts are performed for 5 different radial aperture sizes:         
1.5, 3, 5, 7.5, and 10 arcmin radius with no magnitude limit for the galaxies
selected. Since an aperture size of about $0.5 h_{50}^{-1}$ Mpc in
physical scale corresponding to about two core radii of a rich cluster
provides a good sampling of the high signal-to-noise part of the
galaxy overdensity in a cluster, the chosen set of apertures gives a good
redshift coverage in the range from about $z = 0.02$ to $0.3$ as shown
by the values given in Table 4. With this choice and the depth limit
of the COSMOS data set we are aiming at a high completeness in the
cluster search out to a redshift of about $z = 0.3$. For this goal
the chosen flux limit and the depth of the COSMOS data base are quite
well matched as the richest and most massive clusters are still 
detected in both data sets out to this redshift.

\begin{figure}                                                                  
\psfig{figure=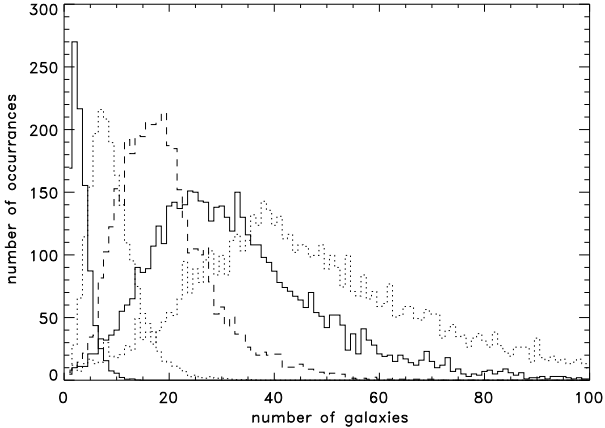,height=6.0cm}                                     
\caption{Example of the distributions of galaxy number counts,
 $\phi (N_{gal}) \times N_{count}$,                   
for the five apertures with radii of 1.5, 3, 5, 7.5, and 10 arcmin.
The histograms are constructed from counts at 1000 random positions
per photographic plate and the results for each aperture size as shown
here are obtained from an average of five plates.
The second, third, forth, and fifth histograms have been multiplied by 
factors of 2, 4, 5, and 6, respectively, for easier comparison.}\label{fig8} 
\end{figure}

\begin{figure}
\psfig{figure=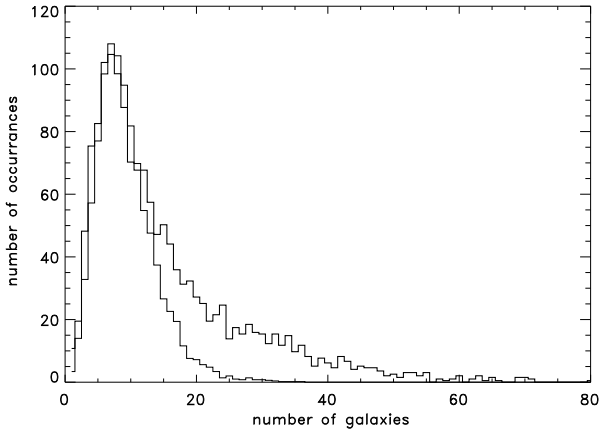,height=6.0cm}
\caption{Example of the distributions of galaxy number counts
in a circular aperture with 3 arcmin radius for an average of five 
UK Schmidt plates and 1000 random positions per plate (thin line).
This distribution is compared to the results of the galaxy number counts for
the 4206 X-ray sources of the sample for the same aperture radius.
The histogram for the random position counts has been normalized
to the histogram of the X-ray source counts so that the peaks have the
same hight.}\label{fig9}   
\end{figure}

The galaxy counts around the given             
X-ray source positions are compared with the number count distributions         
for 1000 random positions for each photographic plate. 
With this comparison we are also accounting for plate to plate variations
in depth as explained below. The number count         
histograms for the random positions have been generated at the                
Naval Research Laboratory in preparation of a COSMOS galaxy cluster catalogue,
the SGP pilot study (Yentis et al. 1992, Cruddace et al. 2000), and for         
this ESO key program. The results of the random          
counts yield a differential probability density distribution, 
$\phi (N_{gal})$, of finding a number    
of $N_{gal}$ galaxies at random positions. An example for the distribution
$\phi (N_{gal})$ for an average of 5 randomly selected plates is shown in 
Fig.~\ref{fig8} for all five aperture sizes. (Note that $\phi (N_{gal})$
is defined here as a normalized probability density distribution
function while in Fig.~\ref{fig8} we show histograms of the form
 $\phi (N_{gal}) \times N_{count}$).  
The distribution functions resemble
Poisson distributions (The possible theoretical description of the functions
is not further pursued here since we are only interested in the purely 
empirical application to the following statistical analysis). In Fig.~\ref{fig9}    
the random count histogram for aperture 2 (3 arcmin radius) is 
compared to the counting results for the 4206 X-ray source positions.
We note the large number of sources with significant galaxy
overdensities in the X-ray source sample compared to the random counts,
and expect to find the  X-ray clusters in this high count tail
of the distribution.

These results for  $\phi (N_{gal})$ are then used          
in the form of cumulative probability distribution functions                  

\begin{equation}                                                      
P( < N_{gal}) = \int_0^{N_{gal}} \phi (N'_{gal}) dN'_{gal} 
\end{equation}                
                                                                
to assign the probability value $P( < N_{gal})$ to each counting result.
                                                                                
For the counts around X-ray sources we expect a significant galaxy density      
enhancement for those sources which have cluster counterparts.                  
Therefore the counting results for the X-ray source positions should             
yield a distribution function $\phi (N_{gal})_X$ which has a more pronounced 
tail at high values of $N_{gal}$ (Fig.~\ref{fig9}). Instead of characterizing the 
enhancement of the counts at high galaxy numbers in the tail of 
$\phi (N_{gal})_X$ we use another data representation as follows.

Going back to the random sample, taking each of the values of 
$P( < N_{gal})$ assigned to each counting result, and plotting
the distribution function $\phi (P( < N_{gal})) \equiv \phi (P(N))$
we will find that this function is a constant. This follows
simply from the chain rule of differentiation in the following
way

\begin{equation}
\phi (P(N))  dP = \phi (N_{gal})~ \left| {dN_{gal}  \over dP} \right| ~ dP
\end{equation}
 
\begin{equation}
= \phi (N_{gal}) \left({d\over dN_{gal}}~ \int_0^{N_{gal}} 
\phi (N'_{gal}) dN'_{gal})
\right)^{-1} ~dP = const.    
\end{equation}

\begin{figure}                                                                  
\psfig{figure=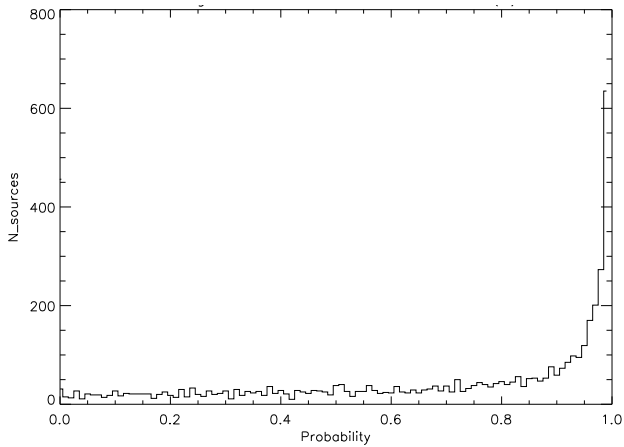,height=6.0cm}                                        
\caption{Histogram of the galaxy excess probabilities, $P_X$, obtained         
from galaxy counts in circular apertures of radius 3 arcmin for the             
sample of 4410 sources of our count rate limited sample of southern             
RASS II sources. There is a clear excess of counts at high values                
of the galaxy density (high values of $P_X$), which is primarily due            
to the effect of galaxy cluster counterparts to the X-ray sources. }\label{fig10}
\end{figure}

Thus for random counts we should expect to see a constant function 
(with noise if the counts are derived in an experiment independent 
from the random count experiment used to define  $P( < N_{gal})$). 
In the case of counts around X-ray sources involving
clusters of galaxies the function $\phi (P_X(N))$ is no longer a 
constant but should show an enhancement for large values of 
$P_X(N)$. Fig.~\ref{fig10} shows the resulting distribution function for
the galaxy counts, $\phi (P_X(N)) \times N_{sources}$,
in the 3 arcmin ring aperture for the sample
of 4410 X-ray sources. The enhancement at large $P$ values
is very pronounced. 

For the further evaluation of this type of diagrams we make the following
simplifying assumptions: i) the distribution function is composed of
two types of counting results, results obtained for cluster
X-ray sources and results obtained for other sources, and ii) the
non-cluster X-ray sources are not correlated to the galaxy
distribution in the COSMOS data base and thus constitute effectively
a set of random counts. This latter assumption is of course not strictly
true for all the non-cluster X-ray sources. While it may be justified to 
treat stars and other galactic sources as well as distant quasars
as independent of the nearby galaxy distribution, there is also a population
of extragalactic sources like low redshift AGN and starburst-galaxies
that we know are correlated to the 
large-scale structure in the galaxy distribution. 
However, the practical assumption that this correlation 
is weak in comparison to the galaxy density enhancements in clusters of 
galaxies is generally well justified. 

\begin{figure}                                                                 
\psfig{figure=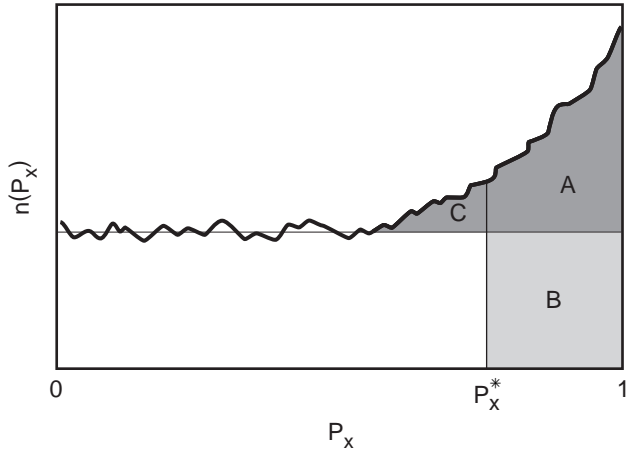,height=6.0cm}                           
\caption{Sketch of the typical result of the distribution function
$P_X(N_{gal})$ for an X-ray source sample containing galaxy cluster
counterparts. The parameter $P^{\star}_X$ indicates the minimal 
allowed value of $P_X(N_{gal})$ in the sample selection. $A + C$
gives the ``true number of clusters'' and $A + B$ the size of
the extracted sample. 
}\label{fig11}
\end{figure}

With this assumption we expect to find a distribution function
$\phi (P_X(N))$ composed of a constant function and a peak at
high P-values. Subtracting the constant function 
leaves us with the cluster sources. This is 
schematically illustrated in Fig.~\ref{fig11}.
For the selection of the cluster candidates we can now either select
the sources which feature a high value of $N_{gal}$ or a high
value of $P_X(N)$. We choose to use  $P_X(N)$ for the
sample selection (as justified further below) in such a way that most
of the cluster peak is included in the extracted sample (that is
choosing  $P_X(N)$ such that the fraction $C$ in Fig.~\ref{fig11}
of cluster lost from the sample is small or negligible).

The clear distinction between the flat distribution for probability             
values between 0 and about 0.7 and the clear and prominent ``cluster peak''
as found in Fig.~\ref{fig10}         
indicates that we can quantify this result further. As illustrated in            
the sketch of Fig.~\ref{fig11}, the cluster contribution is responsible 
for the dark shaded areas labled $A$ and $C$. 
Extracting a sample highly enriched in clusters                 
by choosing a particular high value, $P^{\star}_X$, leaves us
with a formal completeness of the sample expressed by
                                                                                
\begin{equation}
F_{comp} = {A \over A + C}                                       
\end{equation}

The formal contamination of this sample by non-cluster sources 
is likewise given by the expression
                              
\begin{equation}                                                  
F_{cont} = {B \over A + B}                                         
\end{equation}                

The reason for choosing the parameter $P_X(N)$ for the selection of the
cluster sample has also the following reason. The distribution
$\phi (N_{gal})$ is computed for each plate. Since there are plate 
to plate variations in the average galaxy density, using just
$N_{gal}$ would introduce a bias in the sample 
extraction. The use of the parameter $P_X(N)$ takes these variations 
into account. Possible
variations in the background density of the galaxies within each
plate are not accounted for in this approach. 
{\it This is quite a 
general problem, however, to decide at which scale these variations are
taken into account. We expect this residual bias to be small and to
be compensated by the sampling of an overabundance of cluster 
candidates in our strategy to obtain a high completeness.}

The analysis was carried out for all five circular apertures.                 
The strategy for the selection of the cut value, $P^{\star}_X$,                  
was to roughly obtain a sample with 90\% completeness for the                    
single ring statistics and a contamination not much larger                      
than about 20\% to 30\%. Comparing the results for different apertures,
one notes that the peak is best defined for the counts with the two             
smallest apertures. With increasing aperture size the peaks get               
broader and broader, leading to a more and more unfavorable value               
for completeness versus contamination result. Therefore we have                  
relaxed the completeness criterion for the three largest apertures              
to values lower than 90\% not to increase the sample contamination         
dramatically. The resulting values for $P^{\star}_X$, $F_{comp}$,                
$F_{cont}$, and the resulting sample sizes are given in Table 4
for each aperture counting result. We also indicate in the Table
the ``sample size'', defined as $A + B$, and
the ``number of true clusters'' given by $A + C$ (in Fig.~\ref{fig11}).
Note that the sample size is larger than half of the starting sample       
(4410 objects) and that the results indicate the presence of roughly  
1800 galaxy clusters, a number much larger than expected.
                                                                                
   \begin{table*}                                                                
      \caption{Statistics of the results of the galaxy counts around            
the 4410 X-ray sources (above the first count rate cut). 
Columns 2 to 5 give the lower probability limit
for the sample selection, the sample completeness, the contamination,
the sample size and the estimated total number of clusters (see
text for more details). Column 6 gives the
physical scale of the aperture radius at a redshift $z = 0.08$,
close to the median redshift of the REFLEX sample. Column 7 gives
the redshift at which the aperture radius corresponds to a physical
size of $0.5 h_{50}^{-1}$ Mpc. The combined sample is defined by
all candidates flagged in at least one of the aperture count searches.
}
         \label{Tempx}                                                          
      \[                                                                        
         \begin{array}{llrlrrrr}                                             
            \hline                                                              
            \noalign{\smallskip}                                                
 {\rm aperture~ radius}& P^{\star}_X  & F_{comp}  & F_{cont} &              
{\rm sample~ size} & {\rm ``true~ clusters''} & {\rm radius~ (Mpc)} &
{{\rm redshift~ for~ r=0.5~ MPC}}  \\
            \noalign{\smallskip}                                                
            \hline                                                              
            \noalign{\smallskip}                                                
1.5 {\rm arcmin}  & 0.84  & 90\%   & 19\% & 2003 & 1808 & 0.19 & 0.3 \\
3.0 {\rm arcmin}  & 0.84  & 90\%   & 19\% & 2146 & 1930 & 0.37 & 0.12 \\
5.0 {\rm arcmin}  & 0.90  & 75\%   & 18\% & 1582 & 1725 & 0.61 & 0.06 \\
7.5 {\rm arcmin}  & 0.91  & 60\%   & 23\% & 1224 & 1568 & 0.92 & 0.045 \\
10. {\rm arcmin}  & 0.91  & 50\%   & 30\% & 1027 & 1432 & 1.23 & 0.03  \\
{\rm comb. sample}    &    -  & >90\%  &  -   & 2640 & \sim 1750\\                  
            \noalign{\smallskip}                                                
            \hline                                                              
         \end{array}                                                            
      \]                                                                        
   \end{table*}                                                                  
%
%
                                                                                

\section{Inspection of the first selected candidate sample}        
                                                                                
A major reason for the large number of cluster candidates found in the           
above described selection process is easily found by an inspection              
of the optical images of the selected candidates.
The main contribution of spurious clusters comes from bright stars            
($m_b < 12$ mag) and nearby galaxies. For the bright stars the 
diffraction spikes visible on the optical plates are often split
up by the object detection algorithm of COSMOS into a string of single
objects mainly classified as galaxies. Therefore these bright stars appear 
in the statistics as ``clusters of galaxies''. 
Similarly some nearby galaxies
are split up into multiple objects. Both cases are trivially recognized 
in a first inspection of the optical fields around the X-ray sources on 
the POSS and UK Schmidt plates. They can therefore be easily removed from the
sample. Another class of sources enhancing the cluster peak in the  
statistics given in Table 4 are multiple 
detections of very extended X-ray cluster sources.
In these sources the multiple detections have larger separations
than two arcmin (see section 4) 
generally because the center determination settled on
local maxima or photon density fluctuations.
They are also easy to remove by an inspection of the photon distribution 
in the source fields. We have removed redundant detections for all sources
were multiple detections occured within a well connected diffuse source
photon distribution.

In total 452 stars with clearly visible diffraction spikes, 32
nearby galaxies, and 204 redundant detections of diffuse X-ray
sources were removed from the sample. With the cleaned sample 
we can now repeat the statistical analysis with 
results given in Table 5. The sample selection cut
is kept the same as above. (Note that in this statistics there are
about 8\% of the sources missing which leads to a lower normalization
but has no effect on the conclusions drawn in the following).  
We note that this time the statistics indicates a number 
of about $800 - 900$ for the true
clusters in the sample which is close to our
expectations. We also note, that for the combined sample of
candidates from the different aperture sizes,
we obtain a total sample size which is about a factor of $ \sim 1.5 $
larger than the estimated number of true clusters. Thus we expect 
a level of contamination of non-cluster sources 
around $30 - 40\%$. This implies a
laborious further identification work to clean the sample from 
the contamination, a price to be paid for the high completeness level
aimed for.

   \begin{table*}                                                                
      \caption{Statistics of the results of the galaxy counts around            
the X-ray sources in the sample cleaned from bright stars and multiple 
detections. The columns are as in table 4.}
         \label{Tempx}                                                          
      \[                                                                        
         \begin{array}{llllrr}                                                  
            \hline                                                              
            \noalign{\smallskip}                                                
 {\rm aperture~ radius}& P^{\star}_X  & F_{comp}  & F_{cont} &              
{\rm sample~ size} & {\rm ``true~ clusters''}  \\                                   
            \noalign{\smallskip}                                                
            \hline                                                              
            \noalign{\smallskip}                                                
1.5 {\rm arcmin}  & 0.84 & 88.5\% & 22\% & 974 & 863 \\                      
3.0 {\rm arcmin}  & 0.84 & 90\%   & 24\% & 890 & 836 \\                      
5.0 {\rm arcmin}  & 0.90 & 82\%   & 21\% & 703 & 767 \\                      
7.5 {\rm arcmin}  & 0.91 & 65\%   & 26\% & 606 & 718 \\                      
10. {\rm arcmin}  & 0.91 & 52\%   & 32\% & 562 & 690 \\                      
{\rm comb. sample}    &    -  & >90\%  & 30 - 40\%   & 1240 & \sim 850\\
            \noalign{\smallskip}                                                
            \hline                                                              
         \end{array}                                                            
      \]                                                                        
   \end{table*}                                                                  
%
%
                                                                                

\subsection{Comparison of the cluster search statistics with the final
results of the REFLEX Survey}

To analyse how well the cluster selection has worked we anticipate
the results of the REFLEX survey and the final identification
of the cluster candidates.
We repeat the statistical analysis including all the sources from
the starting sample with a flux in excess of $3 \cdot 10^{-12}$ erg
s$^{-1}$ cm$^{-2}$, corresponding to the flux limit of the REFLEX sample  
(1417 sources without the multiple detections) except for the 
stars with diffraction spikes, and the nearby galaxies (1169 X-ray sources).
The results of the cluster search for this high-flux sample
(with the same values for the selection parameter, $P_X^{\star}$ as used
before) and the comparison with the final REFLEX
sample is given in Table 6. 

   \begin{table*}                                                                
      \caption{Cluster search statistics for X-ray sources above a flux
limit of $3\cdot 10^{-12}$ erg s$^{-1}$ cm$^{-2}$ and comparison 
with the final REFLEX sample.  The first 6 columns are as in Table 4. 
Column 7 gives the number
of REFLEX cluster detected by the specific aperture search, column
8 the fraction detected compared to the total REFLEX sample, and Column 9
the contamination fraction found in the candidate sample
selected by the specific aperture.}
         \label{Tempx}                                                          
      \[                                                                        
         \begin{array}{llrrrrrll} 
            \hline                                                              
            \noalign{\smallskip}                                                
 {\rm aperture~ radius}& P^{\star}_X  & F_{comp}  & F_{cont} &              
{\rm sample~ size} & {\rm ``true~ clusters''} & {\rm clusters~ found} &  
F_{det.} & F_{cont^*}\\                                   
            \noalign{\smallskip}                                                
            \hline                                                              
            \noalign{\smallskip}                                                
1.5 {\rm arcmin}  & 0.84 & 91\%  & 18\% & 580 & 530 & 411 & 90\% & 29\%\\
3.0 {\rm arcmin}  & 0.84 & 93\%  & 17\% & 586 & 516 & 433 & 96\% & 26\%\\
5.0 {\rm arcmin}  & 0.90 & 87\%  & 16\% & 487 & 470 & 401 & 89\% & 18\%\\
7.5 {\rm arcmin}  & 0.91 & 75\%  & 19\% & 406 & 448 & 341 & 75\% & 16\%\\
10. {\rm arcmin}  & 0.91 & 64\%  & 22\% & 354 & 438 & 299 & 66\% & 16\%\\
{\rm comb. sample}    &  & \sim 95\%  & 30 - 40\%   & 673 & \sim 500& & & 33\%\\
            \noalign{\smallskip}                                                
            \hline                                                              
         \end{array}                                                            
      \]                                                                        
   \end{table*}                                                                  
%

We note from the results given in Table 6 that the predictions are 
very close to the actual
findings. One has to be careful, however, in the interpretation of this
comparison. In fact the general agreement should not be surprising as we
have used the same statistics to select the sample and we have
not yet used any independent means to include clusters missed by
our search to check the incompleteness independently.
Nevertheless a few results are striking. The number
of clusters predicted to be found is close to the number actually
identified. This shows that the signal observed in the diagnostic 
plots of the type of Fig.~\ref{fig11} is indeed due to galaxy clusters
and there is no large contamination by other objects.
Had we found for example much less clusters
than predicted, we would be forced to speculate on the 
presence of another source population
that mimics clusters in our analysis. This is obviously not the case
and the high $P_X(N)$ signal is correctly representing the clusters
in the REFLEX sample.
Also the trend in the efficiency
of the different apertures in finding the clusters is predicted
roughly correctly. There are only small differences, as for example
that the total number of clusters and the contamination predicted from
the results of the first two apertures are too high and too low,
respectively.

\begin{figure}                                                                 
\psfig{figure=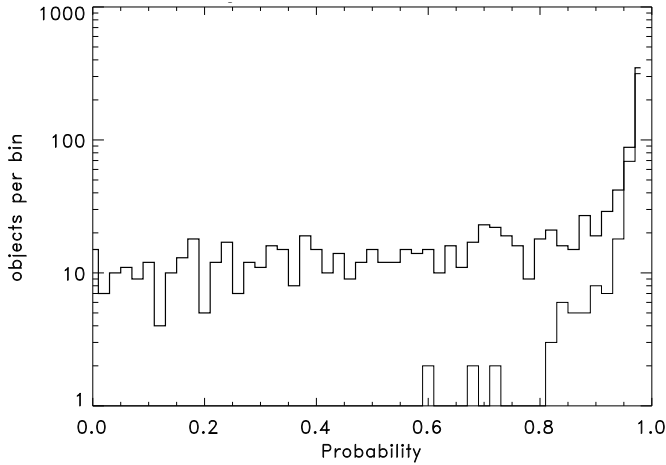,height=6.0cm}                           
\caption{Results for the cluster search with the 3 arcmin aperture radius
for the flux limit of the REFLEX sample. The thick line shows the statistics
for the input sample and the thin line the results for the REFLEX 
clusters. The probability values plotted are defined by equ.(1).
}\label{fig12}
\end{figure}

\begin{figure}                                                                 
\psfig{figure=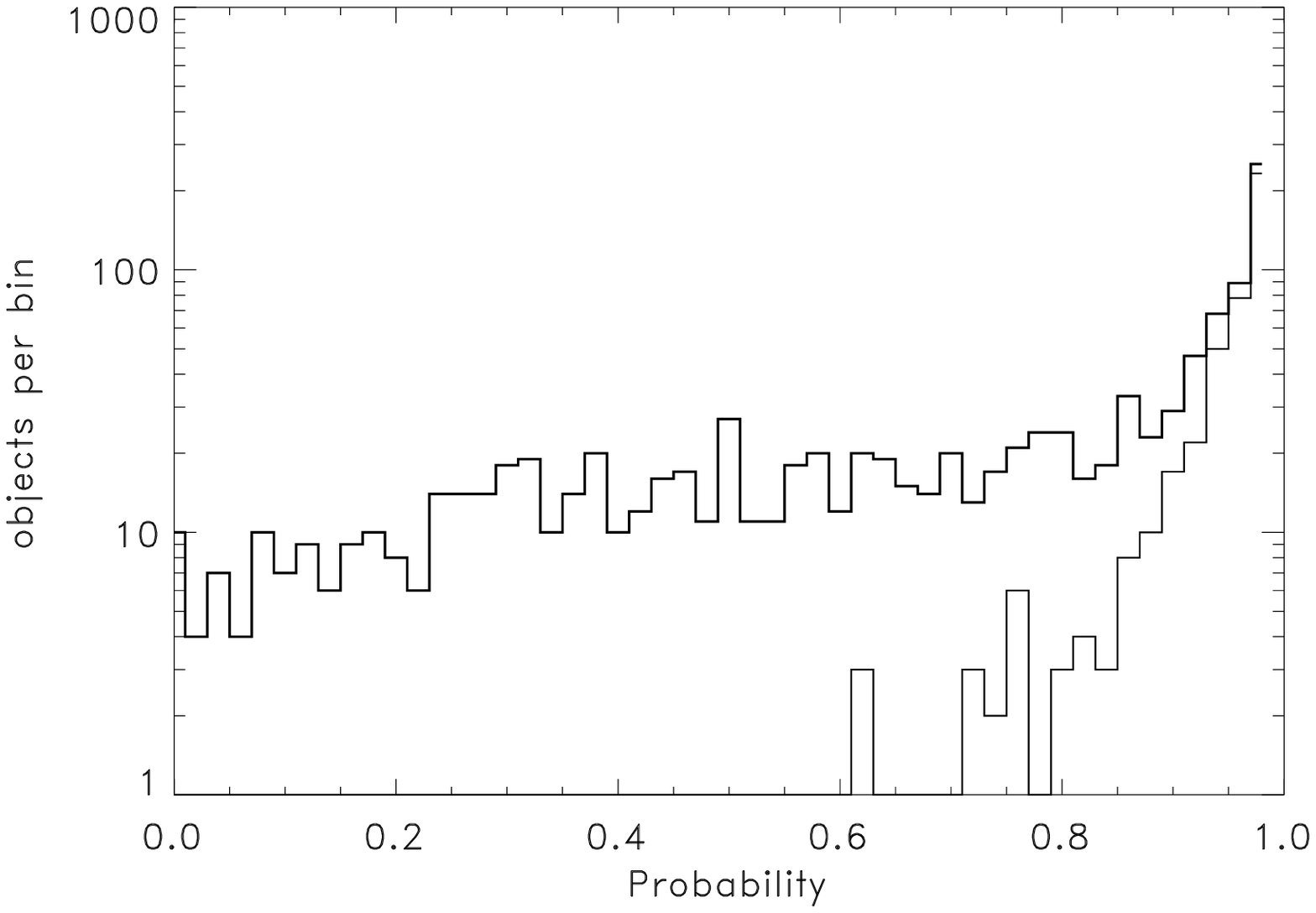,height=6.0cm}                           
\caption{Results for the cluster search with the 5 arcmin aperture radius 
for the flux limit of the REFLEX sample.
The thick line shows the statistics
for the input sample and the thin line the results for the REFLEX 
clusters.}\label{fig13}                                  
\end{figure}

\begin{figure}                                                                 
\psfig{figure=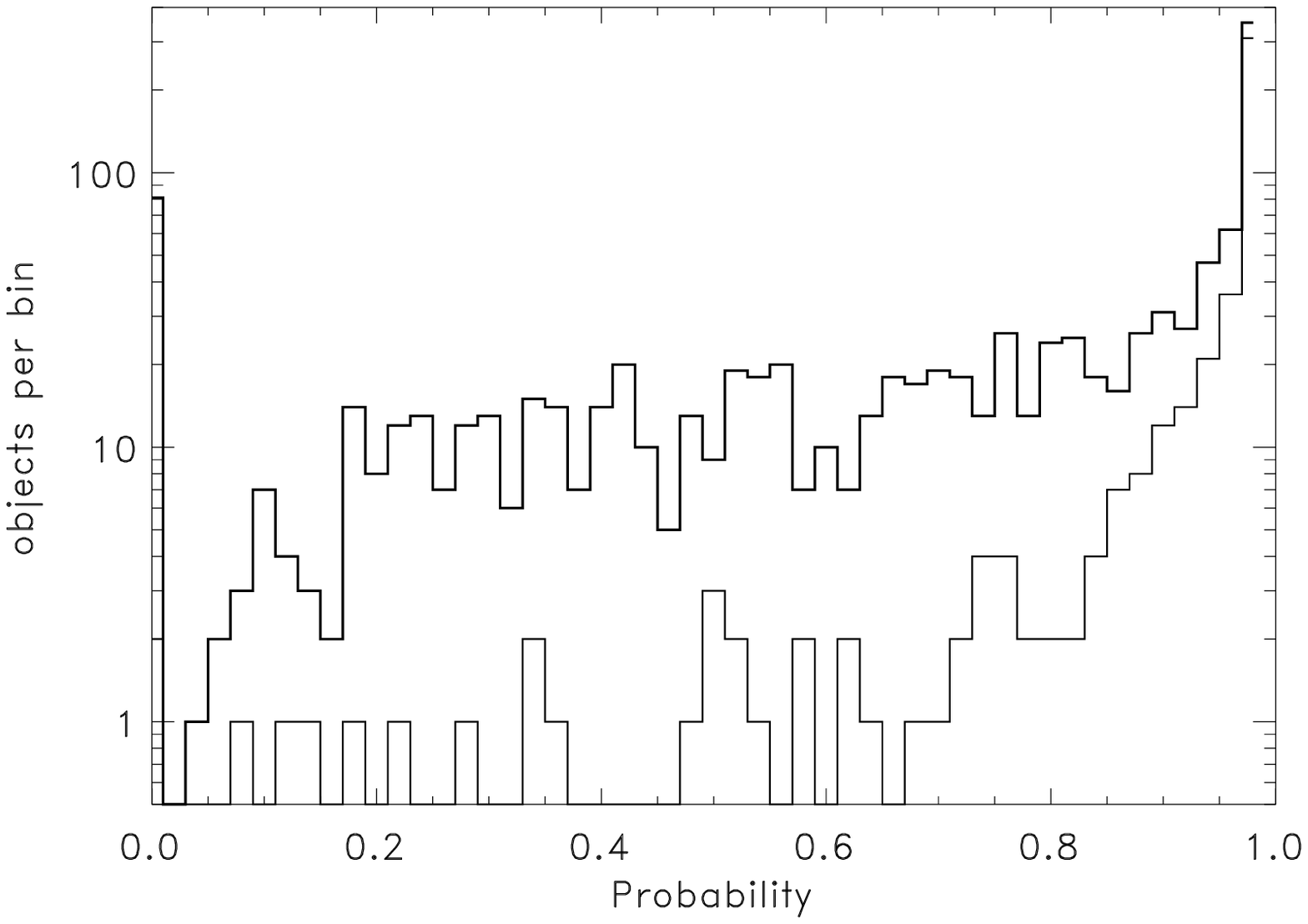,height=6.0cm}                           
\caption{Results for the cluster search with the 1.5 arcmin aperture radius 
for the flux limit of the REFLEX sample.
The thick line shows the statistics
for the input sample and the thin line the results for the REFLEX
clusters.}\label{fig14}                                  
\end{figure}

\begin{figure}                                                                 
\psfig{figure=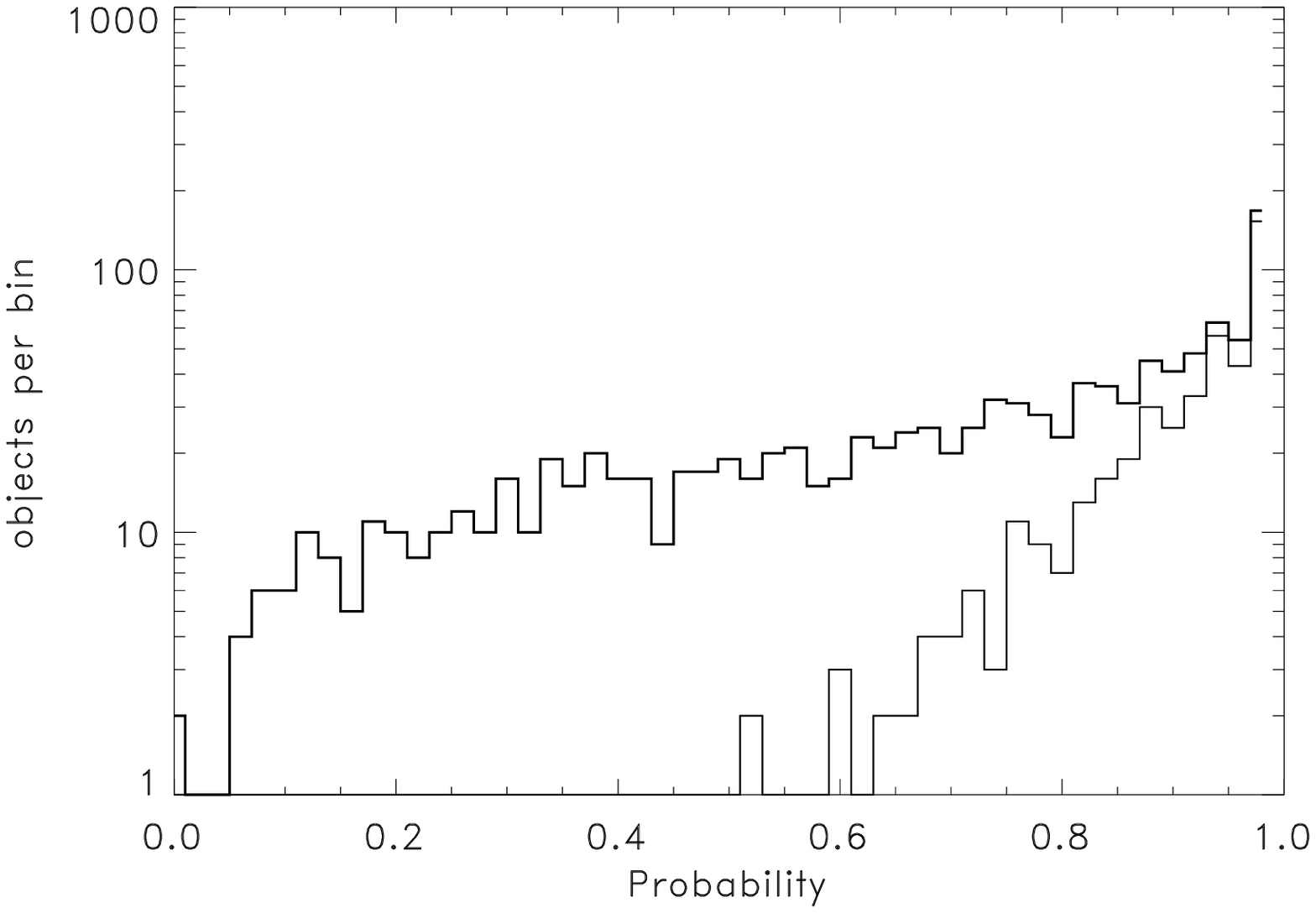,height=6.0cm}                           
\caption{Results for the cluster search with the 10 arcmin aperture radius 
for the flux limit of the REFLEX sample.
The thick line shows the statistics
for the input sample and the thin line the results for the REFLEX
clusters.}\label{fig15}                                  
\end{figure}

\begin{figure}
\psfig{figure=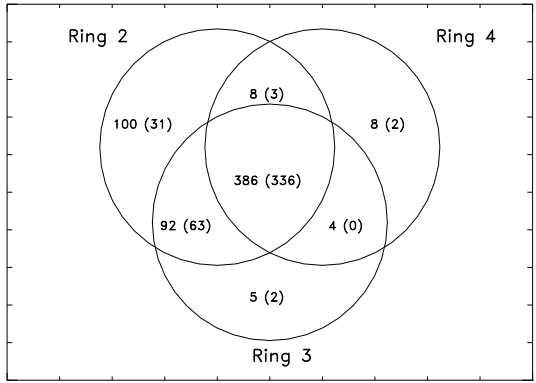,height=6.0cm}
\caption{Number of cluster candidates selected by means of the aperture counts in
rings 1 to 3 (3, 5, 7.5 arcmin, respectively) and number of clusters found in
the REFLEX Survey (values in brackets).}\label{fig16}
\end{figure}

Most of the clusters identified in the REFLEX survey (96\%) 
are detected in the search with aperture 2 with a radius of 3 arcmin. 
That this ring size is the most effective is also shown in Fig.~\ref{fig12} 
where we compare the distribution of the probability values $P_X(N)$ for 
the source sample and for
the subsample which was identified as clusters in the course of
the REFLEX Survey. Only 16 additional clusters are found in aperture
1 with 1.5 arcmin radius and only 3 in the last three apertures.
The peak of the cluster signal is less constrained in the apertures
1, 3, 4, and 5 as shown in Figs.~\ref{fig13} to ~\ref{fig15}. 
The large overlap in the detection of the clusters with the different 
apertures is illustrated in Fig.~\ref{fig16} for 
apertures 2, 3, and 4. Compared to the statistics in the starting sample
the predicted completeness has increased for the high flux sample
mostly due to the fact that the cluster signal in the statistical 
analysis becomes better defined with increasing flux limit. Thus
the statistics of aperture 2 alone gives an internal completeness
estimate of 93\%. Since the results of the different searches
are highly correlated (see Fig.~\ref{fig16}) we cannot easily combine the
results in a statistically strict sense. A rough estimate is given
by a simple extrapolation from the completeness and the sample size
found for aperture 2 and the additional number of cluster found
exclusively in other rings yielding a formal value of 97\%. 

The latter number should be treated with care, however, as an internal
completeness check. This statistics would be more reliable if we had
one homogeneous population of clusters. Since our clusters cover a wide
range of richnesses and redshifts we cannot assume that all subsamples
contribute to the cluster signal in Fig.~\ref{fig11} in the same way. 
If for example
no significant galaxy overdensity could be detected for the high redshift
clusters, this subsample would not enter into the statistics at all.
Likewise, galaxy clusters for which the X-ray detections are missed
in the basic source detection process are also not included in this
prediction.
Therefore the good agreement between the above predictions and 
the final results
supports our confidence in the high quality of the sample but it is not
a sufficient test for completeness. We discuss further 
external tests in section 8.


\begin{figure}
\psfig{figure=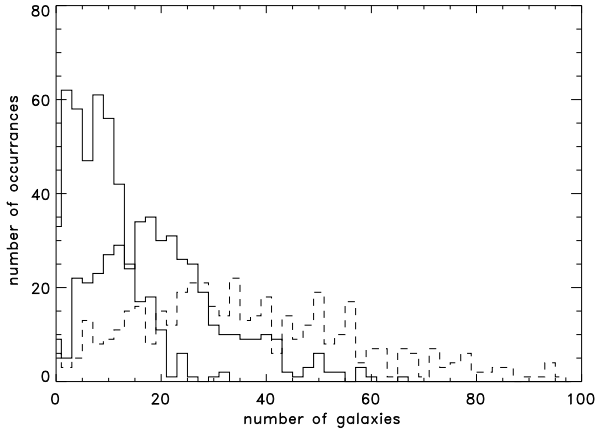,height=6.0cm}
\caption{Distribution of the numbers of galaxies detected in rings 1-3 for
the clusters in the REFLEX sample. Thin line: 1.5 arcmin aperture, thick line:
3 arcmin aperture, broken line: 5 arcmin aperture.}\label{fig17}
\end{figure}
 
\begin{figure}
\psfig{figure=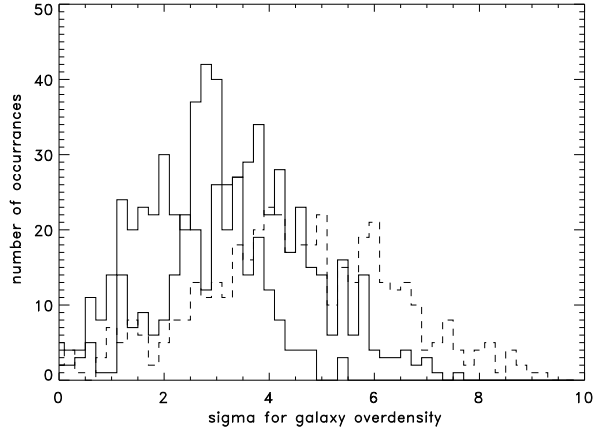,height=6.0cm}
\caption{Distribution of the significance values of the 
detections of galaxy overdensities
in rings 1-3 for the clusters in the REFLEX sample.}\label{fig18}
\end{figure}

Another useful illustration concerns the question of how well defined
the galaxy overdensity signal is for an individual cluster. An answer
is given in Figs.~\ref{fig17} and~\ref{fig18} where 
we show the number
of galaxies (above the background density) for the galaxy
counts in aperture 1, 2, and 3 for the clusters of the REFLEX sample.
For aperture 2 we find for example that the typical count result
is about 20 galaxies per cluster providing a signal of about $4\sigma$.
Thus in general the overdensity signal is very well defined. There
is a tail to low number counts and significances which involves
only a few clusters, however. For aperture 1 we note that the number counts
and significance values are substantially less. Increasing the aperture
size beyond 3 arcmin increases the mean significance of the galaxy
counts, as seen in the results for aperture 3. But what is more
important: the tail towards low significance values is not reduced
if we compare aperture 3 to aperture 2. This once again shows the effectiveness
of aperture 2. One should also note the strong overlap of the results of
the different apertures as illustrated in Fig.~\ref{fig16} which
is reenforcing the significance of the selection results. In fact,
about 60\% of all the REFLEX clusters are flagged in the counting results
for all five apertures. It is not surprising that the second aperture 
with a 3 arcmin radius features as best adapted for our survey, since 3 arcmin 
corresponds to a physical scale of about 370 kpc at the 
median distance of the  REFLEX clusters, $z \sim 0.08$, (see also Table 4).
This corresponds to about 1.5 core radii,
a good sample radius to capture the high surface density part of the clusters.

\begin{figure}
\psfig{figure=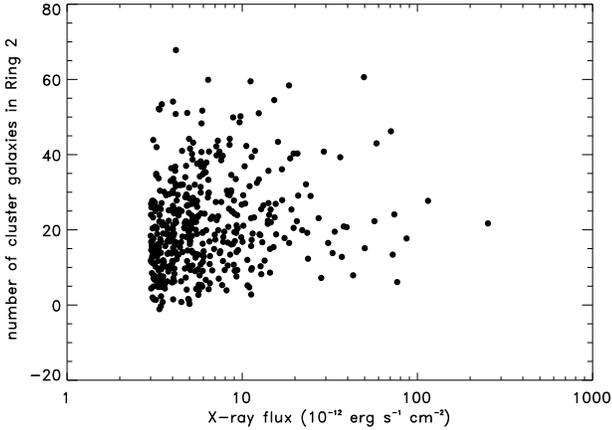,height=6.0cm}
\caption{Distribution of the number of galaxies counted in aperture 2
versus X-ray flux. The background galaxy density has been subtracted
from the aperture counts.}\label{fig19}
\end{figure}

\begin{figure}
\psfig{figure=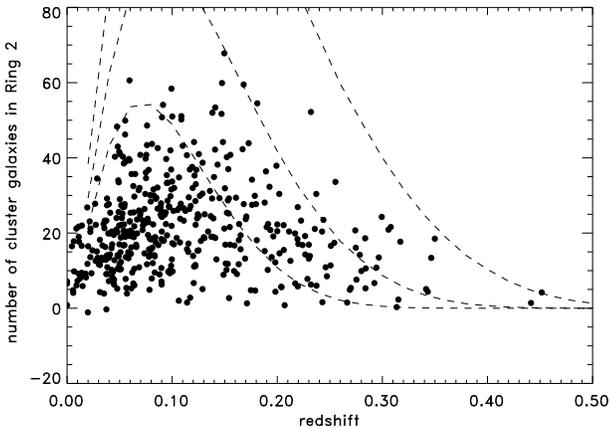,height=6.0cm}
\caption{Distribution of the number of galaxies counted in aperture 2
versus redshift. Also shown are the expected number counts for this aperture
for a cluster with an Abell richness of 100 for the optical magnitude limits
of $b_j = 20, 21, $and$ 22$, respectively. For the cluster model used to
calculate these expected numbers see the description in the text.}\label{fig20}
\end{figure}

\begin{figure}
\psfig{figure=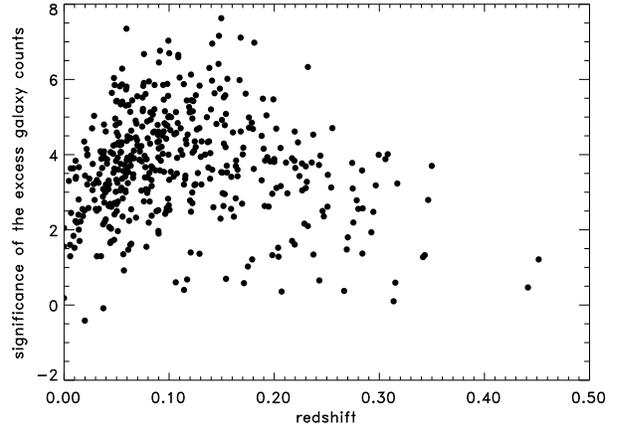,height=6.0cm}
\caption{Distribution of the significance of the galaxy overdensity 
for the galaxies found in aperture 2 versus redshift. Note that the
significance can get negative if the number of galaxies counted in
the aperture is less than the background value.}\label{fig21}
\end{figure}

In Fig.~\ref{fig19} to \ref{fig21} we show the galaxy number 
counts and significances
for aperture 2 as a function of flux
and redshift. While there is no striking correlation with flux we clearly
note the decrease of the number counts and significance values
even for the richest clusters with  redshift. 
In  Fig.~\ref{fig20} we also show the expected number counts for a rich 
cluster (with an Abell richness of 100, which is the number of galaxies 
within $r = 3h_{50}^{-1}$ Mpc and a magnitude interval ranging from the third
brightest galaxy to a limit 2 magnitudes deeper) as a function of redshift 
for three magnitude limits for the galaxy detection on the plates of
$b_j$ = 20, 21, 22 mag. Galaxies are still classified in the COSMOS
data base down to 22$^{nd}$ magnitude but the completeness is decreasing
continuously over the magnitude range from $b_j = 20 - 22$. For the calculation
we assume a Schechter function for the galaxy luminosity function with a slope
of $-1.2$ and $M^*(b_j) = -19.5$ a cluster shape characterized by a King model
with a core radius of $0.5 h_{50}^{-1}$ Mpc, and a k-correction of $\Delta b_j = 3z$
(see e.g. Efstathiou et al. 1988, Dalton et al. 1997). The dashed curves in
Fig.~\ref{fig20} give then the number of galaxy counts expected for the various
magnitude limits. We note that the distribution of the data points 
are well described by the theoretical curves with a steep rise at low 
redshift which is due to an increasing part of the cluster being covered by the
aperture and the decrease at high redshift when only the very brightest 
cluster galaxies are detected. We also not the increasing difficulty to 
recognize clusters above a redshift of $z = 0.3$.

As shown in further
tests in following papers on the REFLEX survey (e.g. Schuecker et al.
2000, B\"ohringer et al. 2000b) there is no deficit of X-ray clusters
in the sample out to a redshift $z = 0.3$ and even beyond, indicating
that the most distant clusters in REFLEX are optically rich enough  to
just be captured by the galaxy count technique applied to the COSMOS data.
Even the most distant clusters in the sample, which have independently
been found as extended RASS sources, are detected and selected
by the correlation based on the COSMOS data. The actual significance
value of $1.2\sigma$ for the extreme case of the most distant REFLEX 
cluster at $z=0.45$, RXCJ1347.4-1144, ($2.5\sigma$ for 
aperture 3) and $0.45\sigma$ for the second most distant cluster 
at $z=0.42$
(found in aperture 1 with a $1.4\sigma$ signal) are quite low
for aperture 2, however. Still, the significance in the optimal aperture
is surprisingly good for the high redshifts of these clusters. 

In summary, we conclude that our combined use of X-ray and optical 
data leads to a very
successful selection of cluster candidates without an introduction
of a significant optical bias, and we expect to be over
90\% complete for the chosen X-ray flux limit.


\section{Further classification of the cluster candidates}

After this anticipation of the final results we return to the
sequence of the REFLEX sample construction.
Up to this point we have compiled a sample of cluster candidates 
relying only on machine based algorithms (except for the manual
exclusion of obvious pathological cases like the bright diffraction 
spike stellar images and the multiple detections).
 The only two selection
criteria are the X-ray flux limit for the X-ray sources and a
signal of a galaxy overdensity in the optical data. This sample 
has still an estimated contamination of non-cluster X-ray
sources of 30 - 40 \% (as discussed in the previous section). 
For the identification work that follows
we treat each source individually, compile as much information as
possible, and try to arrive at a safe classification
in each case. The types
of information used are the X-ray properties of the source,
optical images (from the STScI scans of the POSS and UK Schmidt
plates or CCD images if available), and literature information
(including previous X-ray source identifications).

The basic X-ray source parameters used to assess the source properties
are the probability 
for an extension of the X-ray emission (from the Kolmogorov-Smirnov
test mentioned above) and the spectral hardness ratio. The 
hardness ratio and its photon statistical error is compared
to the expected hardness ratio for a thermal cluster spectrum
for a temperature of 5 keV and the absorbing interstellar column
density at the source position (Dickey \& Lockman 1990), for 
details see B\"ohringer et al. (2000a). As a measure of the 
consistency of the observation with X-ray emission from the
intracluster plasma of a cluster, we take the deviation of the
predicted and observed hardness ratio in units of the statistical error
(in $\sigma$ units).   

In a first step we are discarding all sources that can be 
unambiguously classified
as non-cluster objects. As a safe exclusion criterion we have
either accepted a well documented previous identification
or combined at least two quality criteria which exclude the 
identification as a cluster. Thus the following information 
leads to discarding a cluster candidate source:

i) Positive identification as a non-cluster X-ray source in
the literature.

ii) The X-ray source is both well consistent with a point source
and deviates by more than $3\sigma$ from the theoretically 
predicted hardness ratio of an X-ray cluster. In addition we 
find no indication of a cluster in the optical images.

iii) The X-ray source is point-like and there is an  
AGN spectrum observed for a galaxy or a point-like optical object
at the X-ray source position.

iv) A point-like X-ray source coincides with a $b_j < 12$ mag star 
(within a radius of about 30 arcsec) and there is
no cluster visible in the optical image.

A large part of the contamination fraction can be removed from
the source list by use of these criteria. The positive identifications
are given simply by the observation of a clearly extended X-ray source
and a galaxy cluster in the optical images. For all the X-ray sources
for which no clear classifications can be obtained and also for all
clusters that have not been identified previously and for which no redshift
is available, further spectroscopic and if necessary imaging observations
were conducted. In total 431 targets were observed by us within the ESO
key programme for this project (including candidates with X-ray
fluxes below the current flux limit of the REFLEX survey) as well as additional 
targets in related programmes (e.g Cruddace et al. 2000 in preparation).  

The identification strategy in the optical follow-up observations is 
similar to the scheme given above. We either try to establish the
existence of a galaxy cluster as the counterpart of the X-ray source
by securing several coincident galaxy redshifts in the X-ray source field
or by arriving at an alternative identification of the X-ray source
which is in general an AGN. AGN are found for about 10\% of
the sources for which spectra were taken. More details about the 
identification process and the different types of non-cluster sources
found within the ESO key programme will be given in a subsequent
paper which will also provide the object catalogue. Here
we concentrate on discussing the statistics of some X-ray properties
of the cluster and non-cluster X-ray sources in section 10.
                                                                                
\section{Further tests of the sample completeness} 

To further test the completeness of the REFLEX sample we have
conducted two additional searches for clusters. The first search
is based on the X-ray source extent and the second on a systematic
search for X-ray emission from Abell clusters.

In the search for clusters among the extended RASS sources we have
inspected all sources in the flux limited sample
($F_X \ge 3\cdot 10^{-12}$ erg s$^{-1}$ cm$^{-2}$)
which feature a KS-probability less than 0.01 of being point-like
and which are not already included in the REFLEX sample.
In total 48 additional extended sources are found (after removal
of strange detections at exposure edges and fragments of larger
clusters already included in REFLEX). 35 of these sources are
identified with bright stars and QSO and we notice that they 
are often borderline cases concerning the extent significance; another
fraction of these sources feature an extent in the analysis
without deblending because they are close
pairs of point sources with identifications other than clusters.
The remaining objects for which a cluster identification cannot
be ruled out are in total 13 X-ray sources: 8 certain clusters,
2 good looking cluster candidates and 3 fields with no indication for
a galaxy cluster and no obvious other identification. We
are planning further deeper imaging for the latter sources. 
Thus we have found about 10
objects in this search which have been missed in the REFLEX 
compilation. The above result can also be used for another 
interesting and useful statistic. The 35 partly spuriously extended
sources among the non-cluster candidates (plus the 28 extended 
non-cluster sources mentioned in section 10) if compared to an 
initial sample of 1050 sources (above the REFLEX cut with extended 
cluster sources subtracted) implies a failure rate of flagging 
non-extended X-ray sources erroneously as extended of less than 6\%.  

The Abell and ACO catalogues (Abell 1958, Abell, Corwin, \& Olowin 1989)
contain about 5 times as many objects as the REFLEX sample in the study
area. Even so we do not expect a very close match of the two samples,
since e.g. the correlation of X-ray luminosity and optical richness
is quite weak (see e.g. Ebeling et al. 1993), the large
overabundance of Abell clusters provides a good check regarding 
problems in the recognition of clusters by the galaxy count technique based
on COSMOS. To search systematically for X-ray emission from all 
ACO and ACO supplementary clusters we run the GCA algorithm on all
ACO positions allowing for a recentering of the method within a  
radius of 10 arcmin of the input position. We find only one ACO
supplementary cluster that was not flagged by the galaxy counts and
should be included in REFLEX given its GCA flux. 
This cluster was already found in the above discussed additional
cluster search at the positions of the extended RASS
X-ray sources. It happens that this cluster is actually
close to the boundary to the Large Magellanic Cloud which might explain
the deficiency in counted galaxies at this position. 

Since the search for X-ray emission from ACO clusters is independent
of the previous source detection in the RASS II primary source list,
we are not only testing the completeness of the cluster finding by
the optical galaxy counts but also the source detection in the
RASS II standard analysis (Voges et al. 1999). 
Since we find no ACO cluster missing in REFLEX
due to its non-detection in RASS II, we can conclude that missing of 
sources in RASS II is not a significant problem for the completeness
of the REFLEX sample. Such completeness of the primary source
detection will be studied further by simulations of the source detection
efficiency in the RASS data.

In summary, from the available material we find a missing 
fraction of clusters of about 2 - 3\%
in REFLEX which can be recovered as described in 
this section. This small fraction is still well consistent with the 
internal estimate of a completeness of over 90\% and further supports
the quality of the REFLEX sample. Note that the additional cluster 
detections are not integrated into the REFLEX sample to conserve
its homogeneity but will be listed as REFLEX supplementary clusters
in forthcoming catalogue publications.

\section{Properties of the REFLEX cluster sample}              
                                       
After the identification based on the spectroscopic and imaging 
follow-up observations 452 objects were accepted as galaxy
clusters in the catalogue. For three objects of this list there
is no conclusive redshift available yet and two of these three objects
are still classified as candidates which require a final 
confirmation. 

\begin{figure}
\psfig{figure=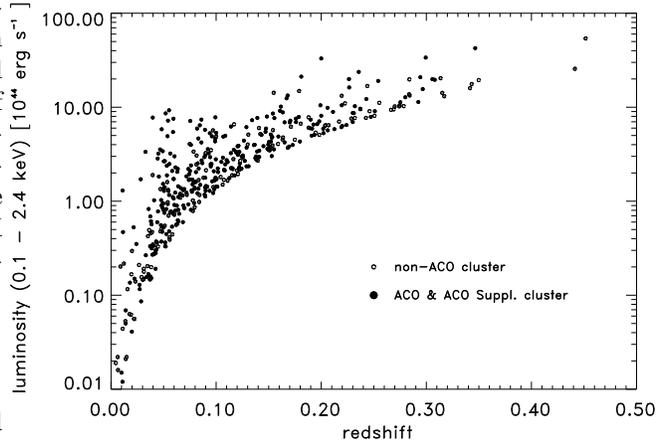,height=6.0cm}
\caption{Distribution of the REFLEX sample clusters in redshift 
and X-ray luminosity. The clusters catalogued by Abell, Corwin \& Olowin
(1989) and the non-ACO clusters are marked differently.}
\label{fig22a}
\end{figure}

Fig.~\ref{fig22a} shows the distribution of the X-ray luminosities and redshifts
for the 449 clusters with redshift information. Details on the way
the fluxes and luminosities of the clusters are calculated  can be
obtained from B\"ohringer et al. (2000a, 2000b). The parabolic boundary
in the plot reflects the flux limit of the sample. The sample is
covering a luminosity range from about $1\cdot 10^{42}$ erg s$^{-1}$
to $6\cdot 10^{45}$ erg s$^{-1}$. The objects with luminosities
below $10^{43}$ erg s$^{-1}$ are Hickson type groups and even
smaller units down to elliptical galaxies with extended X-ray halos.
In the latter objects the extended X-ray emission is still tracing
a massive dark matter halo which is in principle not different 
from a scaled down cluster. Therefore we have included them in the
cluster sample with the caveat that we are not certain at present
how well the population of these objects below a luminosity of 
$10^{43}$ erg s$^{-1}$ is sampled in this project. This is because 
some of them feature a very small membership number which may not always
guarantee that they are detected by the galaxy count search.

At high redshifts, beyond $z = 0.3$, only exceptionally luminous objects
are observed, with X-ray luminosities of several $10^{45}$ erg s$^{-1}$.
Even in this simple distribution plot we can recognize inhomogeneities
in the cluster distribution which can be attributed in a more detailed
analysis to the large-scale structure of the Universe 
(Schuecker et al. 2000, Collins et al. 2000). The sparceness of the data
at very low redshifts in Fig.~\ref{fig22a} is an effect of the small 
sampling volume. The 
apparent deficiency of clusters with $L_x \ge 10^{45}$ erg s$^{-1}$
in the redshift interval $z = 0 - 0.15$ is certainly a cosmic variance 
effect. Only about 3 such X-ray luminous clusters are expected in this
region. While we do not expect the sample to be complete above a redshift
of $z = 0.3$, the expected number of objects at these high redshifts is
indeed very small in a no-evolution model. We explore this further in a 
forthcoming paper.

\begin{figure}
\psfig{figure=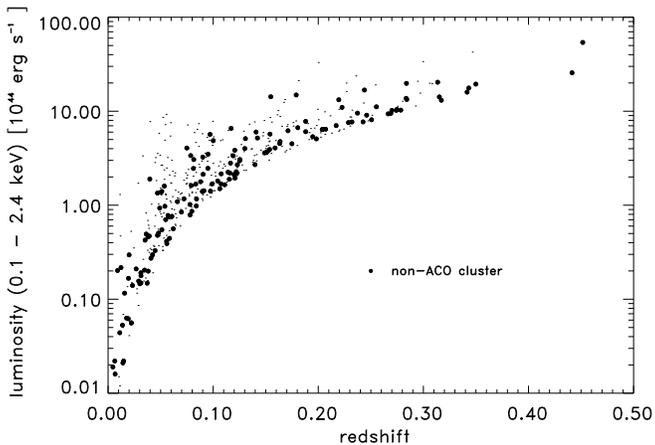,height=6.0cm}
\caption{Distribution of the non-ACO clusters in the REFLEX sample 
in redshift and X-ray luminosity. These clusters cover practically 
the whole distribution range of all  REFLEX clusters.
The clusters catalogued by Abell, Corwin \& Olowin
(1989) are also shown as very light points.}
\label{fig22b}
\end{figure}

Fig.~\ref{fig22a} also shows which of the clusters in the luminosity
redshift distribution are clusters already catalogued by Abell et al. (1989)
and which are mostly new. Since the difference of the two 
different populations is not so easily recognized in this Figure
we have plotted the non-Abell clusters separately in Fig.~\ref{fig22b}.
One notes that the non-ACO clusters are distributed over the whole
range of parameters covered by the total REFLEX sample. As we had
expected, many non-ACO clusters are found among the
nearby low luminosity, poor clusters which fail Abell's richness threshold
and among the most distant clusters, which are not covered well
in the optical plates. To our surprise there is also a large fraction
of non-ACO clusters found in the intermediate redshift range with
X-ray luminosities implying more typical Abell type cluster masses.
These latter clusters indicate an incompleteness effect in the
Abell catalogue.

\begin{figure*}                                                                 
                                                                                
\centerline{\psfig{figure=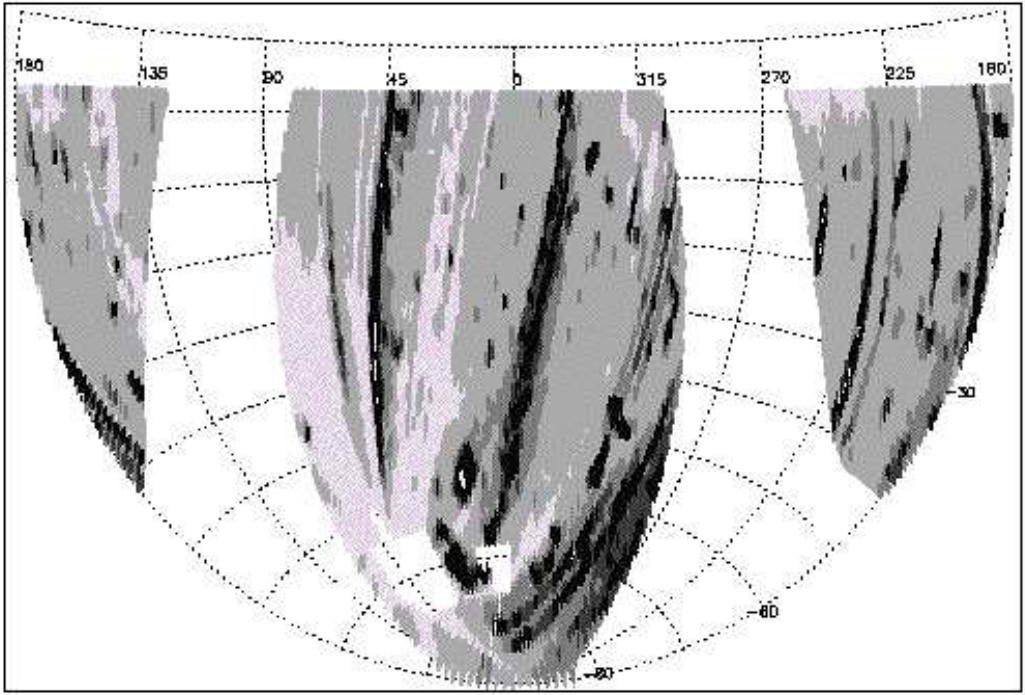,height=9.5cm}}                             
\caption{Sensitivity map of RASS II in the area of the REFLEX survey. 
Five levels of increasing grey scale have been
used for the coding the sensitivity levels given in units of the number of 
photons detected at the flux limit:
$ > 60$ , $30 - 60$ , $20 - 30$, $15 - 20$, and $< 15$,  
respectively.}
\label{fig023}
\end{figure*}

\begin{figure}                                                                 
\psfig{figure=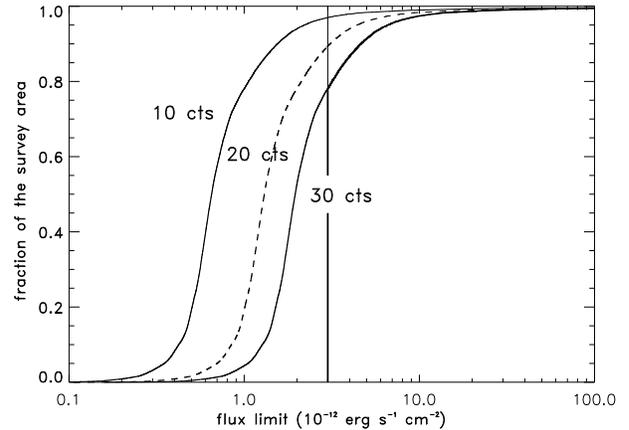,height=6.0cm}                           
\caption{Effective sky coverage
of the REFLEX sample. The thick line gives the effective sky area
for the nominal flux limit of $3 \cdot 10^{-12}$ erg s$^{-1}$ cm$^{-2}$
and a minimum number of 30 photons per source as used e.g. for the correction 
of the $\log$N$-\log$S-curve shown in Fig.~\ref{fig24}. For further details
see text.}\label{fig23}
\end{figure}

Since we do not have a homogeneous exposure coverage of the REFLEX survey
area as described in section 3 we have to apply a corresponding correction
to any statistical study of the REFLEX sample. The best way to take the effect
of the varying exposure and the effect of the interstellar absorption 
into account is to calculate for each sky position the number of photons
needed to reach a certain flux limit. This includes both the exposure
and the sensitivity modification by interstellar extinction. 
In total the sensitivity variation due to extinction is less than
a factor of 1.25 in the REFLEX survey area (see also B\"ohringer
et al. 2000a for details and numerical values). The so defined
sensitivity distribution across the REFLEX study region is shown
in Fig.~\ref{fig023}. 
Since for the relatively 
short exposures in the RASS the source detection process is practically
always source photon limited and not background limited (except for
the most diffuse, low-surface brightness structures) the success rate
of detection depends mostly on the number of photons. 
The use of the ROSAT hard band to characterize the cluster emission
further reduces the background which is a great advantage for this
analysis. Thus fixing a 
minimum number of photons per source we can calculate the effective
survey depth in terms of the flux limit at any position on the sky.
The integral of this survey depth versus sky coverage is shown in
Fig.~\ref{fig23} for the three cases of a minimum detection of 10, 20, and 30
photons. Also shown is the nominal flux limit of $3 \cdot 10^{-12}$
erg s$^{-1}$ cm$^{-2}$. We note that for a detection requirement of
10 photons the sky coverage is 97\% at a flux limit of  $3 \cdot 10^{-12}$
erg s$^{-1}$ cm$^{-2}$. For the much more conservative requirement 
of at least 30 photons per source the sky coverage for the nominal flux
limit of the survey is about 78\%. For the remaining part of the
survey area the flux limit is slightly reduced.
Since the sensitivity map is available for the whole study 
area (Fig.~\ref{fig023}) we can
for any choice of the minimum number of photons calculate the 
correction for the missing sky coverage as a function of flux also
for the three-dimensional analyses e.g. the determination of the
correlation function and the power spectrum of the cluster density
distribution (see Collins et al. 2000, Schuecker et al. 2000).

\begin{figure}
\psfig{figure=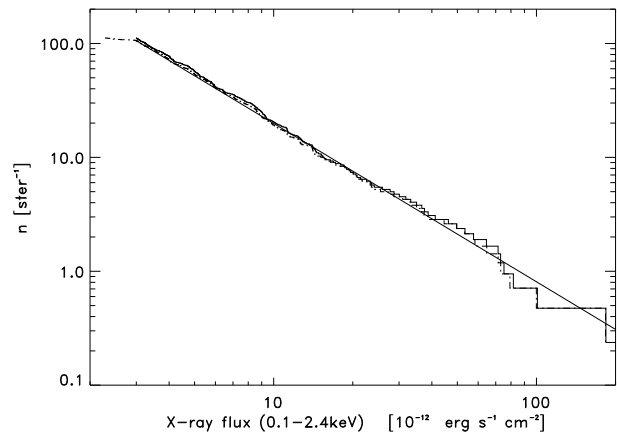,height=6.0cm}
\caption{LogN-LogS-distribution of the REFLEX sample clusters. The dashed
line shows the $\log$N-$\log$S function for the nominal fluxes (determined
for an assumed temperature of 5keV and $z = 0$) which is used for
the REFLEX flux cut while the solid line shows the same function for
the corrected fluxes as described in section 2. The straight line  
shows the result of a maximum liklihood fit of a power law function 
to the data yielding a slope value of $-1.39 (\pm 0.07)$.}\label{fig24}
\end{figure}

In Fig.~\ref{fig24} we give the integral surface number counts of clusters
for the REFLEX sample as a function of X-ray flux 
($\log$N$-\log$S-curve). For this determination we have chosen the 
conservative requirement of a minimum of 30 counts. The Figure also
shows the result of a maximum likelihood fit of a power law function 
to the data for the corrected fluxes. 
The likelihood analysis takes the uncertainties of
the flux measurement (analogous to the description of 
Murdoch et al. 1973) and the variations of the effective
sky coverage for a count limit of 30 photons (as given in Fig.~\ref{fig23})
into account. The resulting power law index is constraint to the range
$-1.39 (\pm 0.07)$. The normalization in Fig.~\ref{fig24} is fixed
to be consistent with the total number of clusters found.
This result is in good 
agreement within the errors with other determinations of the 
cluster number counts as the results
by Ebeling et al. (1998), De Grandi et al. (1999) and Rosati et al. (1998).

\begin{figure}
\psfig{figure=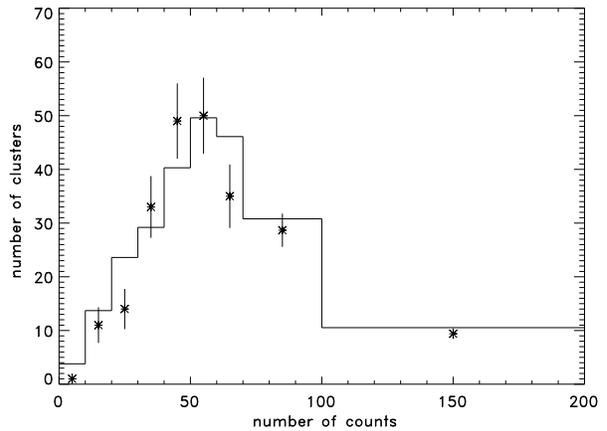,height=6.0cm}
\caption{Distribution of the number of source counts per cluster for the
REFLEX sample. The numbers are given as the number of objects per bin of ten 
photons width. The solid line gives the expected numbers as calculated
from the $\log$N-$\log$S distribution while the stars give the actual
number counts with their Poissonian errors.}\label{fig25}
\end{figure}

Given the $\log$N$-\log$S-distribution corrected for the varying flux limit
as shown in Fig.~\ref{fig24}, we can now also calculate the number of clusters
we expect to be detected with a certain number of counts. This
distribution is shown in Fig.~\ref{fig25}. Here we are first of all
interested in checking the completeness of the sample concerning
detections at low photon numbers ($< 30$ photons). Since the 
$\log$N$-\log$S-distribution was constructed based on clusters with
more than 30 counts only, it provides an independent check on the
relative completeness of the sample for low compared to high photon numbers. 
We note that the number of
clusters to be detected with low photon numbers is quite small and
also that there is no striking deficit of clusters at low counts.
Below a detection with 10 counts 3.8 clusters are expected and
1 is detected. In the interval between a detection of 10 to 20
counts there is no deficit and for the interval between 10 and 30
counts the expectation is about 37 clusters compared to 26 found,
a $2\sigma$ deviation. Therefore we expect very little difference
for the statistical analyses using different cuts in count rate,
as long as the corresponding sky coverage is taken into account.
In fact in the construction of the luminosity function we find
only a difference of less than 2 percent 
(in the fitting parameters for an analysis using a 10 photon count
and a 30 photon count limit, respectively (B\"ohringer et al. 2000b).   
The proper corrections for the effective sky area will become increasingly
important, however, when the sample is extended to lower flux limits.

\section{Statistics of the X-ray properties of the cluster sources
and the sample contamination}              

\begin{figure}[h]
\psfig{figure=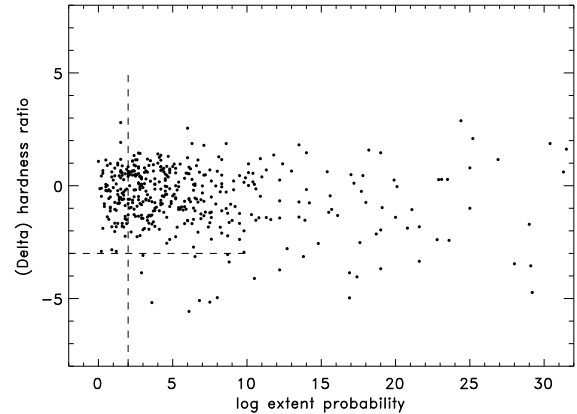,height=6.0cm}
\caption{Hardness ratio deviation and extent probability 
distribution of the REFLEX clusters. The vertical axis
gives the deviation of the measured hardness ratio from the
theoretically calculated value in units of the standard deviation.
For the detailed definition of the parameters and the threshold values
see the text.
}\label{fig26}
\end{figure}

The GCA X-ray source analysis returns two source quality parameters,
the spectral hardness ratio and the source extent. These two parameters
are not used for the selection of the candidate sample. We have
only used this information in conjunction with optical data
as a justification to remove a number of obviously contaminating
sources. Therefore the distribution of these source properties
gives a practically independent information on the nature of the REFLEX cluster
sample and it is interesting to study them in comparison to
the properties of the 
non-cluster sources. In Fig.~\ref{fig26} we show 
the distribution of the two X-ray parameters for the 452
REFLEX clusters. The figure also shows the boundaries used
for the decision as dotted lines: a source is considered to
be very likely extended if it has a Kolmogorov-Smirnov probability 
of less than 0.01 (-$\log$P = 2); and a deviation from the expected 
hardness ratio of more than 3$\sigma$ (to the soft side)
is considered as an 
argument against a cluster identification. Based on these cuts
we find that 81\% of the REFLEX clusters feature an X-ray
source extent. Only 6\% of all the sources have an observed
spectral parameter which appears too soft. This is a small
failure rate which is partly due to statistical fluctuations,
possibily due to an inaccurate acount of the interstellar 
absorption for some of the sources, and also partly due to
the contamination of an AGN in the cluster for some of these
few sources. But since the overall deviation is only significant
for 6\% of the sources the contamination by AGN which might
be indicated here is not a problem for the statistical use
of the overall sample.

\begin{figure}[h]
\psfig{figure=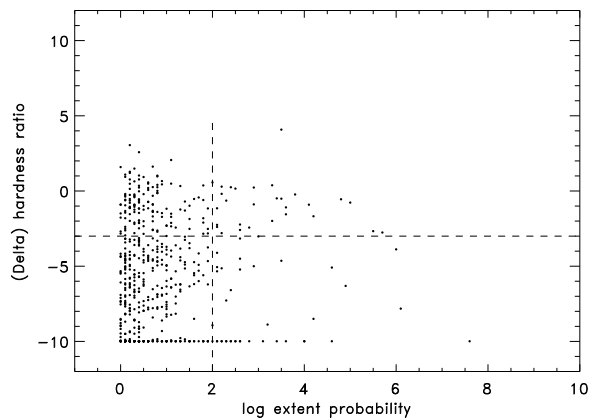,height=6.0cm}
\caption{Hardness ratio deviation and extent probability 
distribution of stars and AGN excluded from the REFLEX 
sample clusters.}\label{fig27}
\end{figure}

It is interesting to compare these source parameters with 
those for non-cluster sources. In Fig.~\ref{fig27} we show the distribution
of the hardness ratio deviations and the source extent probabilities
for the sample of 221 cluster candidates flagged by the galaxy counts
but excluded from the sample in the subsequent identification
process. (Note that this sample has some bias in comparison
to a random non-cluster sample since e.g. (i) in some cases the optical 
selection may be due an extended object falsely split up into galaxies
(ii) contaminating sources may be preferentially recognized if
they have a soft spectrum). 
There is a large fraction of much softer sources. 
About 13\% percent feature an apparent extent, however.
This is more than the failure rate typically found in the 
analysis of a test sample of already identified AGN
which are known point sources and the statistics of the falsely
flagged extended sources shown in section 8 ($< 6\%$). 
The higher rate of detection
of extended sources among these non-cluster sources as compared
to the false classification rate found in Section 8 is partly 
due to really extended 
X-ray emission from nearby galaxies and due to close, blended
double sources. The latter two source types are easily
recognized by inspection and therefore the actual false
classification rate of point sources as extended including
the inspection is at most about half of these 13\%.

\begin{figure}[h]
\psfig{figure=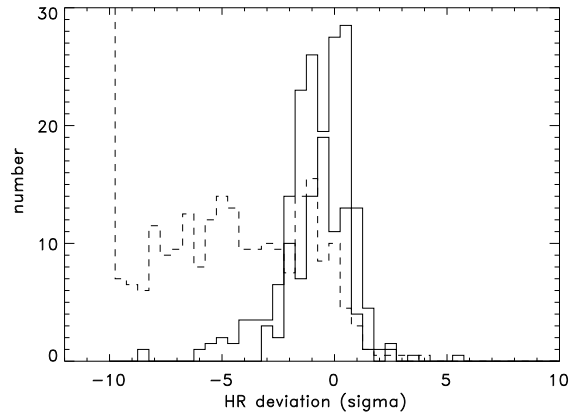,height=6.0cm}
\caption{Distribution of the significance of the deviation
of the measured from the expected hardness ratio: extended
REFLEX clusters (thin line), REFLEX clusters with no significant
extent (thick line), non-cluster sources excluded from the 
REFLEX sample (dashed line). The plot shows that the extended
and non-extended REFLEX clusters do belong to the same spectral
class of X-ray sources.
}\label{fig28}
\end{figure}

We can use the difference in the spectral hardness ratio distribution
of the two samples of cluster and non-cluster sources to test  
for the possible contamination of the REFLEX cluster sample by AGN which 
are producing the dominant X-ray emission in a cluster. First 
of all the high fraction of extended sources guarantees that
the emission of most of these 81\% of the X-ray sources is extended
emission from the intracluster medium of a cluster. The question
is more critical for the non-extended sources. One way of
checking the AGN contamination among them
is to make a statistical comparison between the spectral parameters
of the extended and non-extended cluster sources in REFLEX. This
is done in the form of histograms for the deviation of the measured and 
expected hardness ratio for the two REFLEX subsamples in Fig.~\ref{fig28}.
We note that the two distributions are very similar, quite in
contrast to the very different distribution of 
the non-cluster sources shown in the same figure. Thus there
is no indication that the point-like REFLEX clusters 
are spectrally significantly different from the extended ones.

\begin{figure}[h]
\psfig{figure=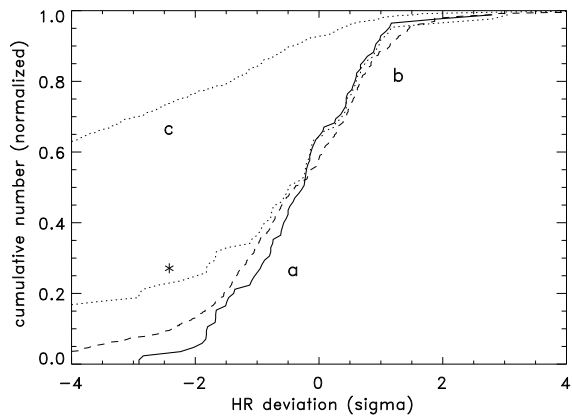,height=6.0cm}
\caption{Cumulative distribution of the significance of the deviation
of the measured from the expected hardness ratio: extended
REFLEX clusters (dashed line, b), REFLEX clusters with no significant
extent (thick line, a), non-cluster sources excluded from the
REFLEX sample (c), non-extended REFLEX clusters artificially 
contaminated by 20 non-cluster sources (*).}\label{fig29}
\end{figure}

This can be more critically tested with cumulative and normalized plots of 
these same distributions as shown in Fig.~\ref{fig29}. Here we note again the 
similarity of the distributions for the two REFLEX subsamples. They have
the same median and the only difference is a slightly broader distribution
for the extended cluster sources. We noted this behaviour already for the
NORAS cluster sample (B\"ohringer 2000a) and it is most probably due to 
the fact that the extended sources contain many more photons on average
and therefore systematic deviations play an increasing role compared
to the pure photon statistics which is the only aspect included in the 
error calculation. The non-cluster sources labled $c$ have a completely
different distribution. To test the sensitivity of this comparison
we have artificially contaminated the point-like X-ray source cluster 
sample by 20 randomly selected non-cluster sources. The resulting 
distribution function is labled with an asterisk in Fig.~\ref{fig29}. This 
sample is significantly different from the cluster distribution 
and such a deviation would easily be recognized. Thus the contamination
in the total sample as introduced by the false identification 
of non-extended REFLEX sources is definitely less than 4\%. 
To this we have to add the possible contamination in the sample
of extended sources which could in principle be due to non-cluster
sources falsely flagged as extended. Making the following very extreme
assumptions: (i) there are as many non-cluster sources as cluster sources
in the candidate sample, (ii) the false classification rate is as high
as 6\% as found in section 8, (iii) all these falsely classified objects
have escaped our careful inspection in the sample cleaning process,
we find an upper limit for the possible contamination of this
part of the sample of less than 5\%. Therefore 
the overall contamination cannot be larger than
9\% and is probably much less.

\section{Summary and conclusions}                                               

The main aim of the construction of this X-ray flux-limited galaxy cluster
sample is its application to measure the large-scale structure of the Universe
and to obtain constraints on cosmological models. 
To this end the sample has to be very homogenous in all its selection
parameters in particular in its coverage of the sky. The unavoidable
inhomogeneities have to be well quantified and modeled.
Here, we described the construction 
of the cluster sample and the selection function (shown in Fig.~\ref{fig023}) and have
given the first demonstration that we have achieved our initial goal.

The primary candidate sample has been constructed from the refined second
analysis of RASS II and we have used a starting list of detections
that includes an overabundance of sources almost down to the $2\sigma$
detection limit. To ensure that we do not introduce a bias against the 
flux measurement of extended sources we have reanalysed the sample with
the GCA method which accounts for this difficulty (see B\"ohringer et al. 
2000a for checks of this method with deeper X-ray observations). Independent 
checks for X-ray cluster sources which might have been missed by the 
source detection in the standard analysis of the RASS 
by means of the Abell catalogues have not shown a single
case of a failed detection. 

The second selection is based on a correlation with the galaxy distribution
in the COSMOS data base. Alternatively we could have used a combined means
of identification of the X-ray sources by correlating also with other galaxy
or cluster catalogues to enhance the findings of clusters. The selection
based on only one criterion was deliberately chosen because it is the best
means to guarantee a fairly homogeneous sampling and to have some control
on possible selection effects which can be tested (e.g. we did not find a signature
in the correlation of the cluster density with the quality of the plate
material - to be shown in a following paper in this series). 
We have actually found an additional small fraction ($\sim 2\%$) of clusters 
which would fulfill the X-ray criteria of the sample as described
in section 8, but they are not part of the REFLEX sample to preserve the 
homogeneity of the present cluster catalogue. The second important point in
the optical selection is the achievement of a high completeness. The smaller the 
missing fraction, the smaller is the imprint of the optical selection criteria
on the overall sample.  With an estimated completeness in excess of 95\%
the imprint should be negligible resulting in an effectively X-ray selected sample
of galaxy clusters. This is another important feature of the catalogue since 
in the following application we will build on the close correlation between 
X-ray luminosity and cluster mass. The high estimated completeness of the catalogue 
is to a large part the result of a substantial oversampling of cluster candidates
in the correlation process as described in sections 5 and 6. 
(At present we like to limit the statement about the high completeness
of the sample to the luminosity range $L_X \ge 10^{43}$ erg s$^{-1}$
and redshifts $z \le 0.3$ until these regimes are explored with further
studies.) The price to be paid for this was
the large contamination fraction by non-cluster sources of about 30 - 40\%,
which required a comprehensive follow-up observation programme.

The following identification work, necessary to remove this substantial 
contamination has to be very rigorous not to introduce an uncontrolled 
bias at this step. Therefore the strategy was adopted that either a clear
identification could be achieved or in the case of a classification by
selection in parameter space at least two strong selection parameters
(failure rate not larger than 10\% for each) were required to rule 
out a cluster identification. 

All these measures taken together are the base of the quality of the 
present sample and its high completeness. There is still the question
as to its contamination. It is for example difficult to rule out in each
case that the cluster contains an AGN which is producing the majority 
of the measured X-ray flux. For this case the standard optical 
identification, to secure several coincident galaxy redshifts to
proof the presence of a cluster, does not help to clear-up the situation.
The high fraction of true source extents that could 
be established by our reanalysis and the further tests based on
the statistics of the spectral parameter distribution 
(section 10) show that this is not a serious problem compromising
the statistical use of the sample.   

We conclude that we have reached the aim of the project to
establish a cluster catalogue which can be used for a variety of cosmological 
studies. Part of these are described in a series of papers already submitted
or in preparation covering further tests and the construction
of the correlation function (Collins et al. 2000, paper II), the power 
spectrum of the cluster density distribution (Schuecker et al. 2000, paper III)
the clustering on very large scales (Guzzo et al. 2000), 
and the X-ray luminosity function (B\"ohringer et al. 2000b).


\begin{acknowledgements}                                                        
We thank Joachim Tr\"umper and the ROSAT team providing the RASS data
fields and the EXSAS software, Rudolf D\"ummler, Harald Ebeling, Andrew
Fabian, Herbert Gursky, Silvano Molendi, Marguerite Pierre, 
Thomas Reiprich, Giampaolo
Vettolani, Waltraut Seitter, and Gianni Zamorani for their help in the
optical follow-up observations at ESO and for their work in the early
phase of the project, and Kathy Romer for providing some unpublished
redshifts. We also thank Daryl Yentis and the team at 
NRL for providing some of the software used in connection with the
analysis of the COSMOS data.  We have in particular benefited from the use
of the COSMOS digitized optical survey of the southern sky, operated
by the Royal Observatory Edinburgh.
This work has made use of the SIMBAD database operated at CDS, Strasbourg,
and of the NASA/IPAC Extragalactic Database (NED) which is operated
by the Jet Propulsion Laboratory, California Institute of Technology under
contract with NASA. 
P.S. acknowledge the support by the Verbundforschung under the grant
No.\,50\,OR\,9708\,35, H.B. the Verbundforschung under the grand
No.\,50\,OR\,93065. L.G. acknowledges financial support by A.S.I.
\end{acknowledgements}

\end{document}